\documentclass[structabstract]{aa}

\newcommand{\vONE}[1]{#1}

\usepackage[fleqn]{amsmath}
\usepackage{graphicx}
\usepackage{grffile}
\usepackage{subfigure}
\usepackage{txfonts}
\usepackage{natbib}
\bibpunct{(}{)}{;}{a}{}{,}
\usepackage{MyStyle}
\usepackage[pdftex=true,colorlinks=true]{hyperref}
\usepackage[all]{hypcap}	
\hypersetup{
pdfpagemode = {UseOutlines},
baseurl = {http://www.mpia.de/homes/kuiper},
pdftitle = {On the stability of radiation-pressure-dominated cavities},
pdfauthor = {R.~Kuiper},
pdfkeywords = {Stars: winds, outflows - Stars: massive - Stars: formation - Radiative transfer - Hydrodynamics - Methods: numerical} 
}

\title{On the stability of radiation-pressure-dominated cavities}

\author{ 
R.~Kuiper\inst{1,2}
\and H.~Klahr\inst{2} 
\and H.~Beuther\inst{2} 
\and Th.~Henning\inst{2} 
}
\authorrunning{R.~Kuiper et al.}

\institute{ 
Argelander Institute for Astronomy, 
Bonn University,
Auf dem H\"ugel 71, 
D-53121 Bonn, 
Germany \\
\email{kuiper@astro.uni-bonn.de}
\and 
Max Planck Institute for Astronomy, 
K\"onigstuhl 17, 
D-69117 Heidelberg, 
Germany
}

\date{ Received {\it date} / Accepted {\it date} }

\abstract
{
When massive stars exert a radiation pressure onto their environment that is higher than their gravitational attraction (super-Eddington condition), they launch a radiation-pressure-driven outflow, which creates cleared cavities.
These cavities should prevent any further accretion onto the star from the direction of the bubble, although it has been claimed that a radiative Rayleigh-Taylor instability should lead to the collapse of the outflow cavity and foster the growth of massive stars. 
}
{
We investigate the stability of idealized radiation-pressure-dominated cavities, focusing on its dependence on the radiation transport approach used in numerical simulations for the stellar radiation feedback.
}
{
We compare two different methods for stellar radiation feedback: gray flux-limited diffusion (FLD) and ray-tracing (RT).
Both methods are implemented in our self-gravity radiation hydrodynamics simulations for various initial density structures of the collapsing clouds, eventually forming massive stars. 
We also derive simple analytical models to support our findings.
}
{
Both methods lead to the launch of a radiation-pressure-dominated outflow cavity. 
However, only the FLD cases lead to prominent instability in the cavity shell. 
The RT cases do not show such instability; once the outflow has started, it precedes continuously.
The FLD cases display extended epochs of marginal Eddington equilibrium in the cavity shell, making them prone to the radiative Rayleigh-Taylor instability.
In the RT cases, the radiation pressure exceeds gravity by 1-2 orders of magnitude.
The radiative Rayleigh-Taylor instability is then consequently suppressed. 
It is a fundamental property of the gray FLD method to neglect the stellar radiation temperature at the location of absorption and thus to underestimate the opacity at the location of the cavity shell.
}
{
Treating the stellar irradiation in the gray FLD approximation underestimates the radiative forces acting on the cavity shell.
This can lead artificially to situations that are affected by the radiative Rayleigh-Taylor instability.
The proper treatment of direct stellar irradiation by massive stars is crucial for the stability of radiation-pressure-dominated cavities.
}


\begin{document}

\maketitle

\section{Introduction}
\label{sect:Introduction}
Outflows are a unique characteristic of the star formation process in general.
The strong radiation pressure provided by massive luminous stars plays an important role in the dynamics of these outflows.
\vONE{Observations show that}
massive star-forming regions involve several large-scale outflows and their interaction with the interstellar material depicts an important part of the complex morphology of the cluster center 
\citep[e.g.][]{
Shepherd:1996p18055,
Zhang:2001p18103, 
Beltran:2006p17686, 
Zhang:2007p8480, 
Fallscheer:2009p12914, 
Beuther:2010p13579, 
Quanz:2010p15750}.
Non-radiative MHD simulations \citep{Banerjee:2006p13404, Banerjee:2007p13399, Hennebelle:2011p17895} have shown that cavities of lower density potentially form 
\vONE{before}
the onset of star formation during the collapse of magnetized clouds.
The treatment of the radiation field is the next step in investigating -- especially the thermodynamics of -- the cavities.

From the perspective of numerical development, the implementation and improvement of radiation transport schemes in (magneto-)hydrodynamics simulations is both a challenging and timely problem in modern astrophysics, especially 
in the field of (massive) star formation, in which the radiative feedback onto the environment plays a crucial role \citep{Yorke:2002p1, Krumholz:2007p1380, Krumholz:2009p10975, Price:2009p18113, Commercon:2010p15256, Bate:2010p18153, Kuiper:2010p17191, Kuiper:2011p17433}.

The hybrid radiation transport scheme applied in this paper was previously used in our studies 
of the radiation pressure barrier in the formation of massive stars \citep{Kuiper:2010p17191}
and 
the angular momentum transport in disks around massive stars \citep{Kuiper:2011p17433}.
We introduced the derivation of the solving algorithm as well as its numerical implementation into a three-dimensional (3D) (magneto-)hydrodynamics code in \citet{Kuiper:2010p12874}.
\vONE{The solver algorithm superposes two radiation fields: a frequency-dependent ray-tracing (RT) component representing the stellar irradiation field and a frequency-averaged (gray) flux-limited diffusion (FLD) component representing the thermal dust emission.}
These splitting schemes were previously used  
in one-dimensional (1D) simulations \citep{Wolfire:1986p562, Wolfire:1987p539, Edgar:2003p6}
\vONE{and}
in two-dimensional (2D) hydrostatic disk atmosphere models \citep{Murray:1994p9750} .

The RT methods are commonly used in radiative transfer models without hydrodynamical motion \citep[e.g.][]{Efstathiou:1990p2140, Steinacker:2003p18295}.
Alternative methods for these kinds of radiative transfer models include the
short characteristics method \citep{Dullemond:2000p2185}
and Monte Carlo radiative transfer \citep{Bjorkman:2001p2187, Dullemond:2004p17642}. 
Comparison benchmark studies of these methods were presented in \citet{Pascucci:2004p39} and \citet{Pinte:2009p12529}.
These methods are able to solve the radiation transport problem to high accuracy for a fixed static configuration, but 
\vONE{they}
require much CPU time, which yields low efficiency in time-dependent hydrodynamics simulations. 

In this sense, the FLD approximation \citep{Kley:1989p2162, Bodenheimer:1990p2169, Klahr:1999p963, Klahr:2006p962} provides a fast method to determine the temperature evolution 
\vONE{accurately}
in the optically thick (diffusion) 
\vONE{and}
optically thin (free-floating) limit.
\vONE{However,}
the approximation 
\vONE{becomes inaccurate}
in multi-dimensional anisotropic problems as well as in transition regions from optically thin to thick \citep[shown e.g.~in][]{Boley:2007p2959}.
\vONE{One example of this transition region is}
the cavity shell investigated in this paper.
The FLD approximation in the frequency-averaged (gray) limit remains the default technique in modern multi-dimensional radiation hydrodynamics.
In an exceptional case, a frequency-dependent FLD solver was used in \citet{Yorke:2002p1}.

From the theoretical point of view, the stability of radiation-pressure-dominated cavities has a direct
\vONE{impact on the star formation efficiency and accretion onto a luminous massive star.}
\citet{Krumholz:2009p10975} proposed that massive stars can grow beyond their Eddington limit 
\vONE{because}
these radiation-pressure-dominated cavity shells are subject to the so-called ``radiative Rayleigh-Taylor instability''
\vONE{allowing additional mass to be fed onto the central accretion disk.}
Following the approach of \citet{Nakano:1989p1267} and \citet{Yorke:2002p1}, we proposed the feeding of a massive star beyond its Eddington limit by disk accretion only \citep{Kuiper:2010p17191, Kuiper:2011p17433}. 
While the anisotropy of the thermal radiation field sufficiently diminishes the radiation pressure onto the disk accretion flow near the midplane, the radiation pressure in the polar direction is able to launch an outflow, forming a stable cavity.

The removal of a substantial fraction of the initial core mass by the radiative launching of outflows influences the star formation efficiency in the central core region.
These feedback effects are proposed to play an important role in explaining the low star-formation efficiencies observed 
\citep{McKee:2007p848}.
Simulations of HII regions in massive star forming regions by \citet{Dale:2011p17748} and close to massive stars forming in the cluster center by \citet{Peters:2010p12919} suggest that ionized gas fills preexisting voids and bubbles. 
Long-lived cleared polar cavities therefore would 
\vONE{make} 
the expansion of HII regions and their interaction with the stellar cluster environment 
\vONE{easier}. 
The feedback of the energy injection from jets and outflows onto the stellar cluster formation is briefly discussed in \citet{Bonnell:2006p13823}.

In this study, we investigate different implementation methods for direct stellar irradiation feedback (see Sect.~\ref{sect:Method}) and their influence on the stability of radiation pressure dominated cavities. 
We present in Sect.~\ref{sect:QualitativeResults} the qualitative outcome and Sects.~\ref{sect:QuantitativeResults1} and \ref{sect:QuantitativeResults2} the quantitative analyses of the simulations. 
The observed difference in the radiative acceleration of the cavity shell depending on the applied radiation transport method is analytically derived in Sect.~\ref{sect:Analytic}.
Finally, a comprehensive comparison section (Sect.~\ref{sect:Comparisons}) to previous simulations, analytic work, and observations as well as a brief summary (Sect.~\ref{sect:Summary}) are provided.

\section{Physical and numerical model}
\label{sect:Limitations}
\subsection{Physics included}
We compare the effect of two different radiation transport methods on stellar radiation feedback.
The goal is to investigate the stability of the shell 
\vONE{surrounding the}
radiation-pressure-dominated cavity.
We emphasize that in these simulations we do not aim to provide a complete description of an outflow region around a massive star.

\noindent
A list of potential limitations and caveats includes:
\begin{enumerate}
\item The collimation effects by magnetic fields -- which are certainly important for the morphology of the cavity -- are neglected. Recent results of MHD simulations of jet formation are summarized by \citet{Fendt:2011p17650}. \vONE{Potentially, the relative importance of radiative or magnetic forces in jet formation can be distinguished observationally based on morphology: a magnetically launched fast jet would be more collimated than a radiation-pressure-driven wide-angle outflow \citep[see e.g.][and citations therein]{Arce:2007p2541, Pudritz:2007p4896}.}
\item Another example of the influence of magnetic fields, the so-called photon bubble instability \citep{Turner:2007p1126} could potentially diminish the radiation pressure and could therefore alter the morphology of a cavity filled with magnetized gas.
\item Evaporation of dust is included, but otherwise the dust is assumed to be perfectly coupled to the gas.
\item The dust opacity clearly dominates the absorption in any dusty region, but in the dust-free zone around the massive star up to the dust sublimation front the (remaining) gas opacity is set to be constant $\kappa_\mathrm{gas} = 0.01 \mbox{ cm}^2~\mbox{g}^{-1}$. Such a constant gas opacity of course denotes only a first order approximation to the complex molecular line absorption features.
An optically thick gas disk could e.g.~amplify the flashlight effect \citep{Tanaka:2011p2726} and therefore increase the radiative flux into the polar regions.
\item Since the cavity growth is driven by radiation pressure, the quantitative details of the launching process and the growth phase depend on the stellar evolution model as well as the dust model. Future studies will attempt to establish the details of this dependence.
\item The low-density cavity is potentially enriched, but certainly influenced, by a stellar wind from the central massive star, which is neglected herein. Early proto-stellar winds could, e.g., lead to an early formation of rather collimated polar cavities \citep{Cunningham:2011p2721}. The radiation pressure at later epochs in such a configuration \vONE{could} be channeled into pre-existing cavities, leading to smaller opening angles for the radiation pressure dominated cavities.
\item In \citet{Peters:2010p12919, Peters:2011p17381}, ionizing radiation was proposed to dominate the cavity dynamics in regions of massive star formation.
\end{enumerate}

\subsection{Numerical resolution}
\label{sect:Resolution}
To obtain unambiguous results, the relevant physical length scales have to be resolved by the numerical simulations.
Therefore, we discuss these scales with respect to the resolution of our radiation hydrodynamical simulations.
We demonstrate that the simulations resolve all relevant length scales sufficiently.

A study of a radiative Rayleigh-Taylor instability in the shell surrounding the outflow cavity needs to resolve the wavelengths of this instability along the shell discontinuity.
At small wavelengths, the radiative Rayleigh-Taylor instability is likely to be suppressed by diffusion, but at large wavelengths -- comparable to the physical size of the cavity -- the instability occurs.
Quantitatively, the shortest unstable wavelength for instance is determined to be 1000~AU for a 10\Msol and 10000~AU for a 100\Msol star \citep{Jacquet:2011p18452}.
The polar resolution $r \Delta \theta$ of the spherical grid in our simulations grows linearly with the radius and is fixed to $r \Delta \theta \lesssim 0.1 r$.
The grid size along the shell discontinuity is typically a factor of ten smaller than the shortest unstable wavelength.
Hence, the length scale of the radiative Rayleigh-Taylor instability is fully resolved.

\begin{figure}[t]
\begin{center}
\includegraphics[width=0.48\textwidth]{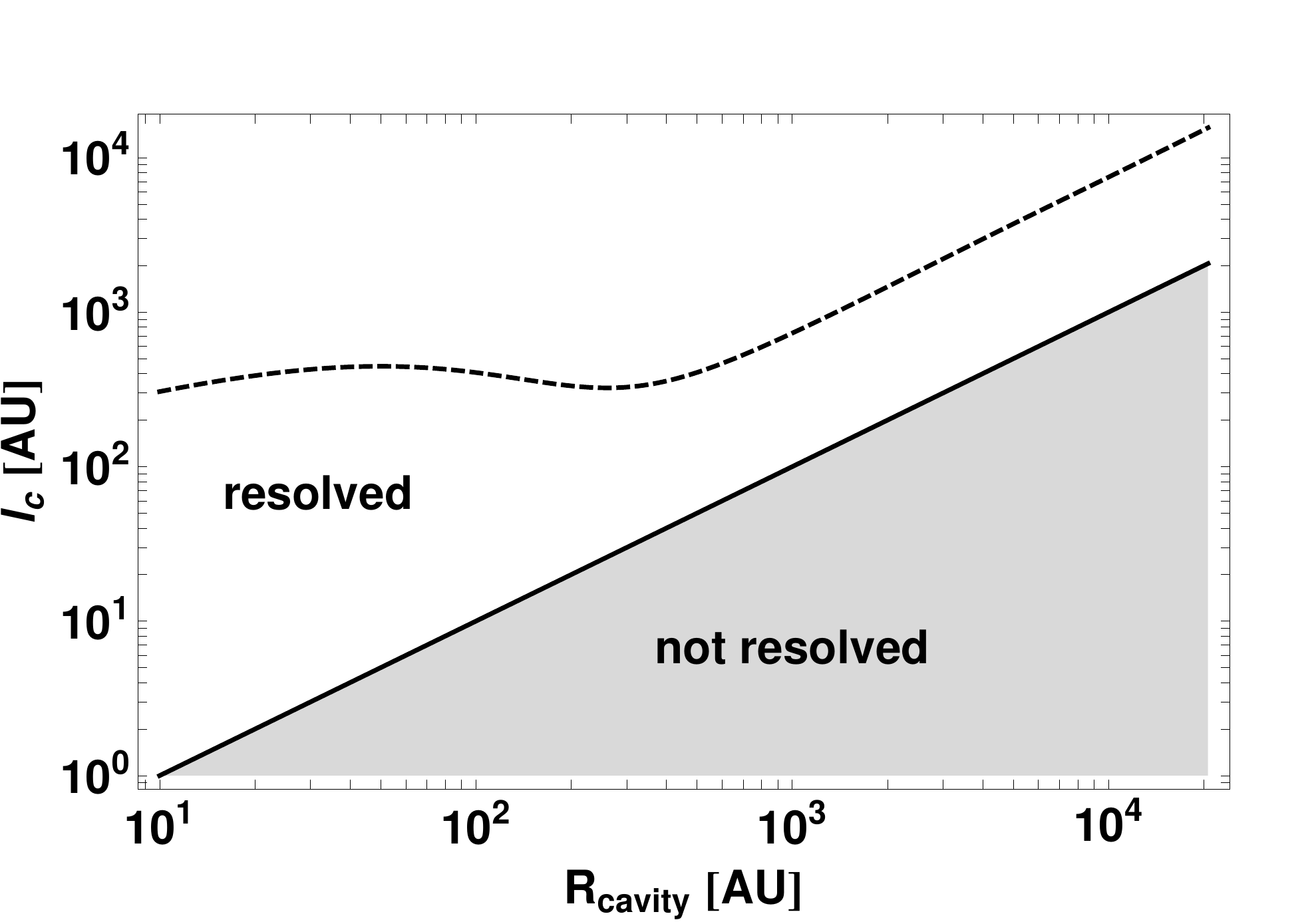}\\
\caption{
The cooling length scale $l_\mathrm{c}$ (dashed line) of the thermal dust emission in the gray approximation as a function of the actual location of the cavity shell $R_\mathrm{cavity}$.
The solid line denotes the resolution of our numerical grid.
}
\label{fig:ResolutionIR}
\end{center}
\end{figure}

Two other important length scales are related to the radiative properties of the shell.
The ``cooling length scale'' $l_\mathrm{c}$ is given by an optical depth of $\tau = 1$ in the long wavelength regime,
and the ``irradiation length scale'' $l_*$ is given by an optical depth of $\tau = 1$ \vONE{of} the broad stellar irradiation spectrum.
Fig.~\ref{fig:ResolutionIR} shows the cooling length scale in the gray approximation as a function of the location of the cavity shell $R_\mathrm{cavity}$.
The shell density is chosen to be $\rho = 4 \times 10^{-17} \rhocgs$ as depicted in Fig.~5.
The opacities are computed as the Rosseland mean opacities of \citet{Laor:1993p736}.
The stellar temperature is taken from the \citet{Hosokawa:2009p12591} tracks for a 20\Msol star and an accretion rate of $\dot{M} = 10^{-3} \Msol \mbox{ yr}^{-1}$, and
the temperature at the location $R_\mathrm{cavity}$ is computed with a slope $T \propto R_\mathrm{cavity}^{-0.5}$ in the gray approximation.
For simplification, the evaporation of dust grains is neglected, but would result in an even larger cooling length-scale.
Hence, we can be sure that the length scale of cooling is resolved.

\begin{figure}[t]
\begin{center}
\includegraphics[width=0.48\textwidth]{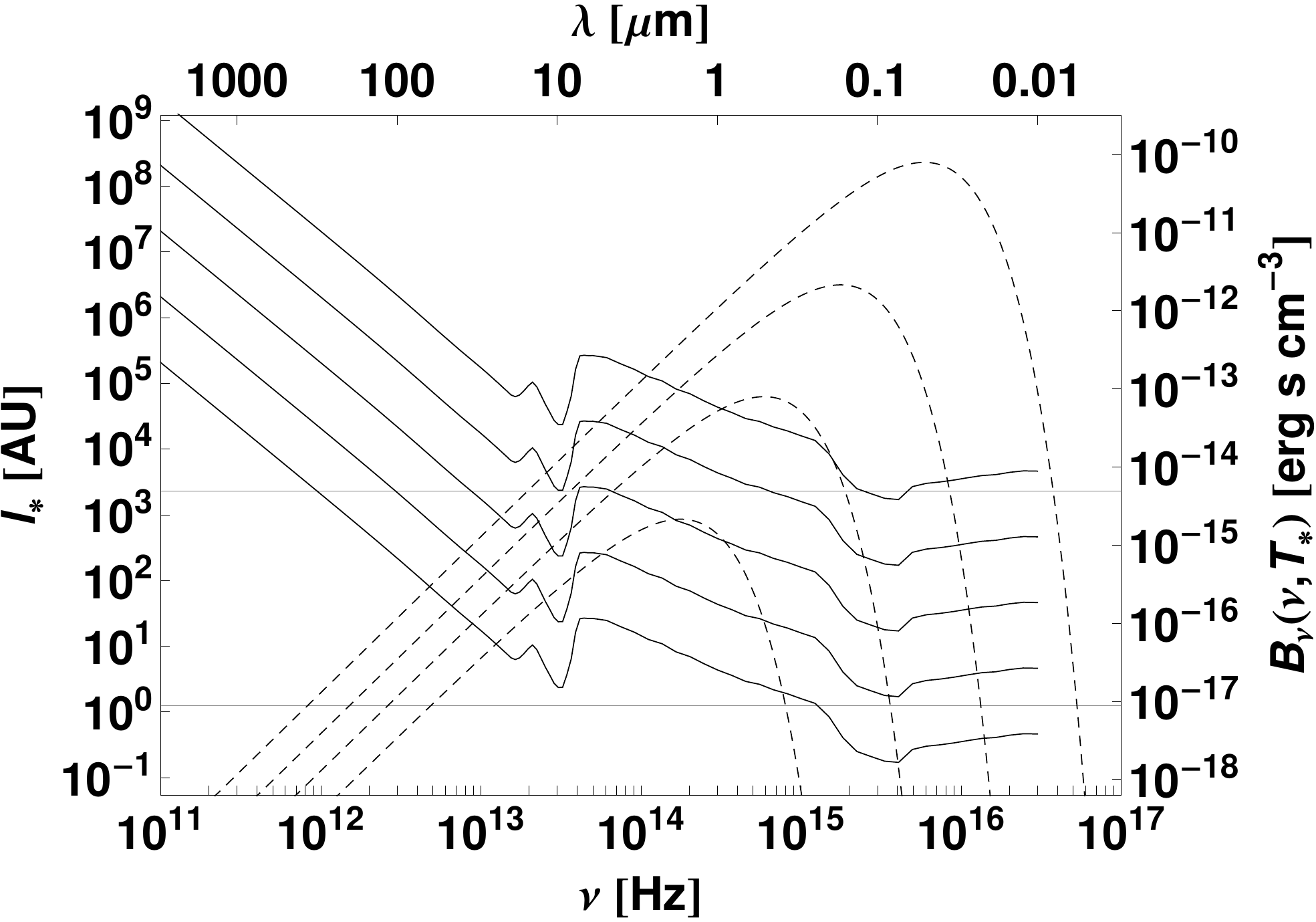}
\caption{
The absorption length scale $l_*$ regarding the stellar irradiation spectrum as a function of frequency $\nu$ or wavelength $\lambda$, respectively.
From top to bottom,
the different solid lines denote a density of $\rho = 10^{-19}, 10^{-18}, 10^{-17}, 10^{-16}, \mbox{ and }10^{-15} \rhocgs$, respectively.
The two horizontal lines mark the highest and lowest resolution of the spherical grid.
The right vertical axis shows the scale for the stellar spectra:
From top to bottom,
the dashed lines denote black body spectra of $T_* = 100000, 30000, 10000, \mbox{ and } 3000$~K.
The effect of potential dust evaporation is not included here, but would lead to a higher resolution of the numerical grid in terms of the irradiation spectrum.
}
\label{fig:Resolution}
\end{center}
\end{figure}

The absorption length scale $l_*$ \vONE{of} the stellar irradiation spectrum is much smaller than the cooling length scale for thermal dust emission.
Using again the opacities of \citet{Laor:1993p736},
Fig.~\ref{fig:Resolution} shows the absorption length scale \vONE{of} the whole stellar irradiation spectrum for different gas densities.
For the aforementioned shell density of $\rho = 4 \times 10^{-17} \rhocgs$ and a typical stellar temperature at the interesting point in time at the potential onset of instability of about $10000$~K~$< T_* < 30000$~K,
the absorption length scale of the photons with frequencies higher than the peak of the stellar black-body spectrum is about 10 to 100~AU; 
these photons are most likely to be absorbed in the first grid cell on top of the cleared cavity.
For lower frequency photons, the absorption length scale smoothly rises up to the 1000~AU scale and is most likely to be resolved.
In the long wavelength regime of the spectrum, the results of the previous paragraph for the thermal dust emission of course apply.

The morphology in the radial direction during the cavity growth phase can be distinguished into three important regimes:
the inner low-density cleared cavity, the shell of swept-up material on top of this cavity, \vONE{and} the remnant of the in-falling envelope on larger scales.
The resolved transition of the gas density, temperature, and velocity in
these regimes is depicted in the context of the following analyses in Figs.~3 (two-dimensional snapshots) and~5 (along the polar axis).
Even the steep gradient of the density at the inner rim of the cavity shell (Fig.~5, upper left panel) is resolved by four grid cells, \vONE{and} the smooth decrease to larger radii by dozens of grid cells.
Therefore, the shell morphology is clearly resolved in the simulations.

\subsection{Axial symmetry}
\label{sect:axialsymmetry}
The simulations in this paper are performed assuming axial symmetry. 
We are convinced that this is not a critical limitation of the result we present. 
In 2D simulations with the same initial conditions, but different methods for stellar radiation feedback, we observe strong differences in the resulting radiation feedback onto the shell on top of the cleared cavity.
These differences do not depend on the detailed morphology of this shell and therefore the conclusion is also applicable to non-axially symmetric, 3D configurations.

In our 2D simulation results with the FLD approximation, the cavity shell undergoes an instability;
this result matches the 3D simulation result of \citet{Krumholz:2009p10975}.
In our 2D simulations as well as in our 3D simulation \citep{Kuiper:2011p17433}, including the RT step for direct irradiation feedback, the radiation-pressure-dominated cavity remains stable.

\section{Method}
\label{sect:Method}
\subsection{Equations}
\label{sect:Equations}
To follow the motion of the gas, we solve the equations of compressible self-gravity hydrodynamics, including shear-viscosity as described in \citet{Kuiper:2010p17191}.
For this task, we use the open source MHD code Pluto \citep{Mignone:2007p3421}.
This set of equations is coupled to the radiation transport in the medium.
In contrast to our previous studies, we investigate the influence of two different methods to determine the radiative flux, namely our hybrid radiation transport scheme (ray-tracing + flux-limited diffusion) and flux-limited diffusion only.

\subsubsection{Ray-tracing + flux-limited diffusion}
\label{sect:RT+FLD}
For simulations labeled ``RT+FLD'', we use our hybrid radiation transport module \citep[see][]{Kuiper:2010p12874}.
The total radiation energy density $\vec{F}_\mathrm{tot}$ is divided into 
the flux $\vec{F}_*(\nu,r)$ of the frequency-dependent stellar irradiation and
the flux of the frequency-averaged thermal radiation energy density $\vec{F}$
\begin{eqnarray}
\label{eq:Radiation_Diffusion1}
\partial_t E_\mathrm{R} + f_\mathrm{c} \vec{\nabla} \cdot \vec{F} &=& - f_\mathrm{c} \left(\vec{\nabla} \cdot \vec{F}_* - Q^+ \right), \\
\label{eq:Radiation_Irradiation}
\vec{F}_*\left(\nu, r\right)/\vec{F}_*\left(\nu, R_*\right) &=& 
\left(R_*/r\right)^2
\exp\left(-\tau\left(\nu, r\right)\right),
\end{eqnarray}
where Eq.~\eqref{eq:Radiation_Diffusion1} denotes the evolution of the thermal radiation energy density $E_\mathrm{R}$.
The factor $f_\mathrm{c} = \left(c_\mathrm{V} \rho / 4 a T^3 + 1 \right)^{-1}$ depends only on the ratio of internal to radiation energy and contains the specific heat capacity, $c_\mathrm{V}$, and the radiation constant, $a$.
The source term $Q^+$ depends on the physics included and contains any additional energy source such as hydrodynamical compression $-P \vec{\nabla} \cdot \vec{u}$ and viscous heating.
We solve Eq.~\eqref{eq:Radiation_Diffusion1} using the so-called flux-limited diffusion approximation, in which the flux is set proportional to the gradient of the radiation energy density ($\vec{F} = - D \vec{\nabla} E_\mathrm{R}$).
The diffusion constant $D = \lambda c / \rho \kappa_\mathrm{R}$ depends on the flux limiter $\lambda$ and the Rosseland mean opacity $\kappa_\mathrm{R}$. 
The quantity $c$ denotes the speed of light in a vacuum.
We use the flux limiter proposed by \citet{Levermore:1981p57} and neglect scattering.

Eq.~\eqref{eq:Radiation_Irradiation} calculates the flux of the frequency-dependent stellar irradiation in a ray-tracing step.
The first factor on the right hand side describes the geometrical decrease in the flux proportional to $r^{-2}$.
The second factor describes the absorption of the stellar light as a function of the optical depth $\tau(\nu,r)=\kappa(\nu) \rho(r) r$ depending on the frequency-dependent mass absorption coefficients $\kappa(\nu)$.
For this purpose, we use tabulated dust opacities by \citet{Laor:1993p736}, including 79 frequency bins, and calculate the local evaporation temperature of the dust grains by using the formula of \citet{Isella:2005p3014}.
The flux at the inner radial boundary is given by the luminosity $L_*$, temperature $T_*$, and radius $R_*$ of the forming star.
For this purpose,  we use tabulated stellar evolutionary tracks for accreting high-mass stars, calculated by \citet{Hosokawa:2009p12591}.
The gas and dust temperature $T$ is finally calculated in equilibrium with the combined stellar irradiation and thermal radiation field
\begin{equation}
a T^4 = E_\mathrm{R} + \frac{\kappa\left(\nu\right)}{\kappa_\mathrm{P}(T)} \frac{|\vec{F}_*|}{c}
\end{equation}
with the Planck mean opacities $\kappa_\mathrm{P}$.

Numerical details, test cases, including a comparison of gray and frequency-dependent irradiation, as well as performance studies of the hybrid radiation transport scheme are summarized in \citet{Kuiper:2010p12874}.
The viscosity prescription as well as the tabulated dust and stellar evolution model are presented in \citet{Kuiper:2010p17191}.

\subsubsection{Flux-limited diffusion only}
For simulations labeled ``FLD'', we do not ray-trace the stellar irradiation, but add the luminosity of the forming star as a source term to the right hand side of the FLD equation
\begin{eqnarray}
\label{eq:Radiation_Diffusion}
\partial_t E_\mathrm{R} + f_\mathrm{c} \vec{\nabla} \cdot \vec{F} &=& f_\mathrm{c} \left(L_* \delta(\vec{r}) + Q^+ \right).
\end{eqnarray}
Since the forming star is enclosed in the inner sink cell in our grid in spherical coordinates, the luminosity of the central star is implemented as a Dirichlet boundary condition at the inner rim of the computational domain.
The radiation energy at this inner boundary is therefore given by the stellar surface temperature and the assumption that the medium between the stellar surface $R_*$ and the inner rim at $r_\mathrm{min} = 10$~AU is optically thin.

The hybrid radiation transport scheme (Sect.~\ref{sect:RT+FLD}) as well as the FLD approach were tested in detail in comparison with a Monte Carlo solution using the RADMC code \citep{Dullemond:2004p17642} or a short characteristics method using the RADICAL code \citep{Dullemond:2000p2185} in a setup including a central star, an accretion disk, \vONE{and} an envelope as described in \citet{Pascucci:2004p39} and \citet{Pinte:2009p12529}.

\subsection{Numerical configuration}
The simulations are performed on a time-independent grid in spherical coordinates with a logarithmically stretched radial coordinate axis.
The outer core radius is fixed to $r_\mathrm{max} = 0.1$~pc, and the inner core radius is fixed to $r_\mathrm{min} = 10$~AU. 
The accurate size of this inner sink cell was determined in a parameter scan presented in \citet{Kuiper:2010p17191}, Sect.~5.1.
The polar angle extends from $0^\circ$ to $90^\circ$ assuming midplane symmetry.
The grid consists of 64 x 16 grid cells, i.e.~the highest resolution of the non-uniform grid is chosen to be 
\begin{equation}
\Delta r \mbox{ x } r\Delta{\theta} = 1.27 \mbox{ AU} \mbox { x } 1.04 \mbox{ AU}
\end{equation}
around the forming massive star.
The resolution decreases logarithmically in the radial outward direction proportional to the radius.
The radially inner and outer boundary of the computational domain are semi-permeable walls, i.e.~the gas is allowed to leave the computational domain (by accretion onto the central star or due to radiative and centrifugal forces over the outer boundary), but cannot enter this domain.
This outer boundary condition allows the mass reservoir for stellar accretion to be controlled by the initial choice of the mass of the pre-stellar core.

The Pluto code uses high-order Godunov solver methods to compute the hydrodynamics, 
i.e.~it uses a shock capturing Riemann solver within a conservative finite volume scheme. 
Our default configuration consists of a Harten-Lax-Van~Leer Riemann solver and a so-called ``minmod'' flux limiter using piecewise linear interpolation and a Runge-Kutta~2 time-integration, also known as the predictor-corrector-method; for comparison we also refer to \citet{vanLeer:1979p5193}. 
Therefore, the total difference scheme is accurate to second order in time and space.

The internal iterations of the implicit solver of the FLD in Eqs.~\eqref{eq:Radiation_Diffusion1} or \eqref{eq:Radiation_Diffusion} is stopped at an accuracy of the resulting temperature distribution of either $\Delta T / T \le 10^{-3}$ or $\Delta T \le 0.1$~K.
The internal iterations of the implicit solver for Poisson's 
equation is stopped at an accuracy of the resulting gravitational potential of $\Delta \Phi / \Phi \le 10^{-5}$.

\subsection{Initial conditions}
\label{sect:InitialConditions}
We start from a cold ($T_0 = 20 \mbox{ K}$) pre-stellar core of gas and dust.
The initially constant dust-to-gas mass ratio is chosen to be $M_\mathrm{dust} / M_\mathrm{gas} = 0.01$.
Dynamically, the model describes a so-called quiescent collapse scenario without initial turbulent motion ($\vec{u}_r = \vec{u}_\theta = 0$)
and
the core is initially in slow solid-body rotation $\left(|\vec{u}_\phi| / R = \Omega_0 = 5*10^{-13} \mbox{ Hz}\right)$.
The total mass $M_\mathrm{core}$ in the computational domain is chosen to be 50 or 100 $\mbox{M}_\odot$.
\vONE{In addition}
to the influence of the different radiation transport methods described in the last sections, we check the dependence of the stability of the radiation pressure dominated cavity on the initial density slope of the pre-stellar core.
We chose the initial density slope to be either $\rho \propto r^{-1.5}$ or $\rho \propto r^{-2}$.
An overview of the runs is given in Table~\ref{tab:runs}.
\begin{table}[htbp]
\begin{center}
\begin{tabular}{l c c c c}
\hline\hline
& & & & \\ 
\raisebox{1.5ex}[-1.5ex]{Run} & \raisebox{1.5ex}[-1.5ex]{$M_\mathrm{core}$} & \raisebox{1.5ex}[-1.5ex]{$\rho \propto r^\beta$} &\raisebox{1.5ex}[-1.5ex]{RTM} &  \raisebox{1.5ex}[-1.5ex]{Epoch of $E \approx 1$}\\ 
\hline
A-RT+FLD	& 100 & -2    & RT+FLD	& No  \\
A-FLD	& 100 & -2    & FLD		& Yes \\
B-RT+FLD	& 50   & -2    & RT+FLD	& No  \\
B-FLD	& 50   & -2	    & FLD		& Yes \\
C-RT+FLD	& 100 & -1.5 & RT+FLD	& No  \\
C-FLD	& 100 & -1.5 & FLD		& Yes \\
\hline
\end{tabular}
\end{center}
\caption{
Overview of simulations performed. 
The table contains 
the run label (column~1), 
the two distinguishing parameters of the different initial conditions, namely the initial pre-stellar core mass $M_\mathrm{core}$ (column~2) and the exponent $\beta$ of the initial density slope $\rho(r)$ (column~3),
in addition to the radiation transport method (RTM) in use (column~4),
and the resulting existence of an epoch of marginal Eddington equilibrium $E \approx 1$ (column~5). 
}
\label{tab:runs}
\end{table}
For a more comprehensive parameter scan of the initial conditions of this pre-stellar core collapse model, we refer to \citet{Kuiper:2010p17191} and \citet{Kuiper:2011p17433}.

\section{Qualitative simulation results}
\label{sect:QualitativeResults}
We summarize the qualitative outcome,
\vONE{the stability,}
and the overall morphology of the radiation pressure dominated cavities, depending on the different initial conditions and the radiation transfer method.
In the six simulations, the radiation pressure is generally high enough to launch an outflow and form a cleared polar cavity.

In the case of the RT+FLD radiation transport method, these outflow cavities, once launched, rapidly grow in their extent up to the outer computational domain at 0.1~pc away from the central star.
In contrast to this monotonic growth, the outflow cavities in the FLD radiation transport scheme stop their increase along the polar axis at an extent of the order of several 100 to 1400~AU depending on the initial conditions.
This stopping of their growth is followed by a penetration of the gas mass on top of the outflow cavity formed so far, i.e. the outflow cavity collapses.
During a follow-up second and third epoch of outflow launching, the resulting cavities grow in their extent, but on much longer timescales than in the RT+FLD case.
The epochs of frozen cavity growth in the FLD-only simulations suggest that (parts of) the top layer of the outflow cavity are in marginal Eddington equilibrium, i.e.~the radiation pressure force is in equilibrium with the stellar gravity.
The qualitative outcome of the simulations (depending on whether an extended epoch of marginal Eddington equilibrium occurs) is summarized in Table~\ref{tab:runs}, column~5.

\begin{figure*}[p]
\begin{center}
\includegraphics[width=0.5\textwidth]{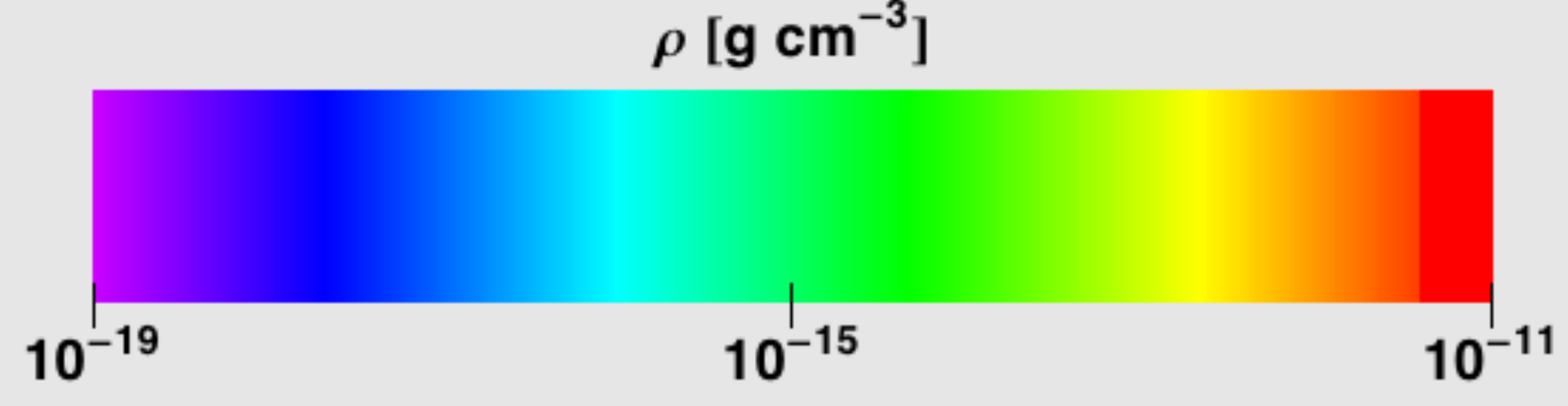}
\end{center}
\subfigure[FLD run at $t = 44$~kyr]{
\includegraphics[width=0.36\textwidth]{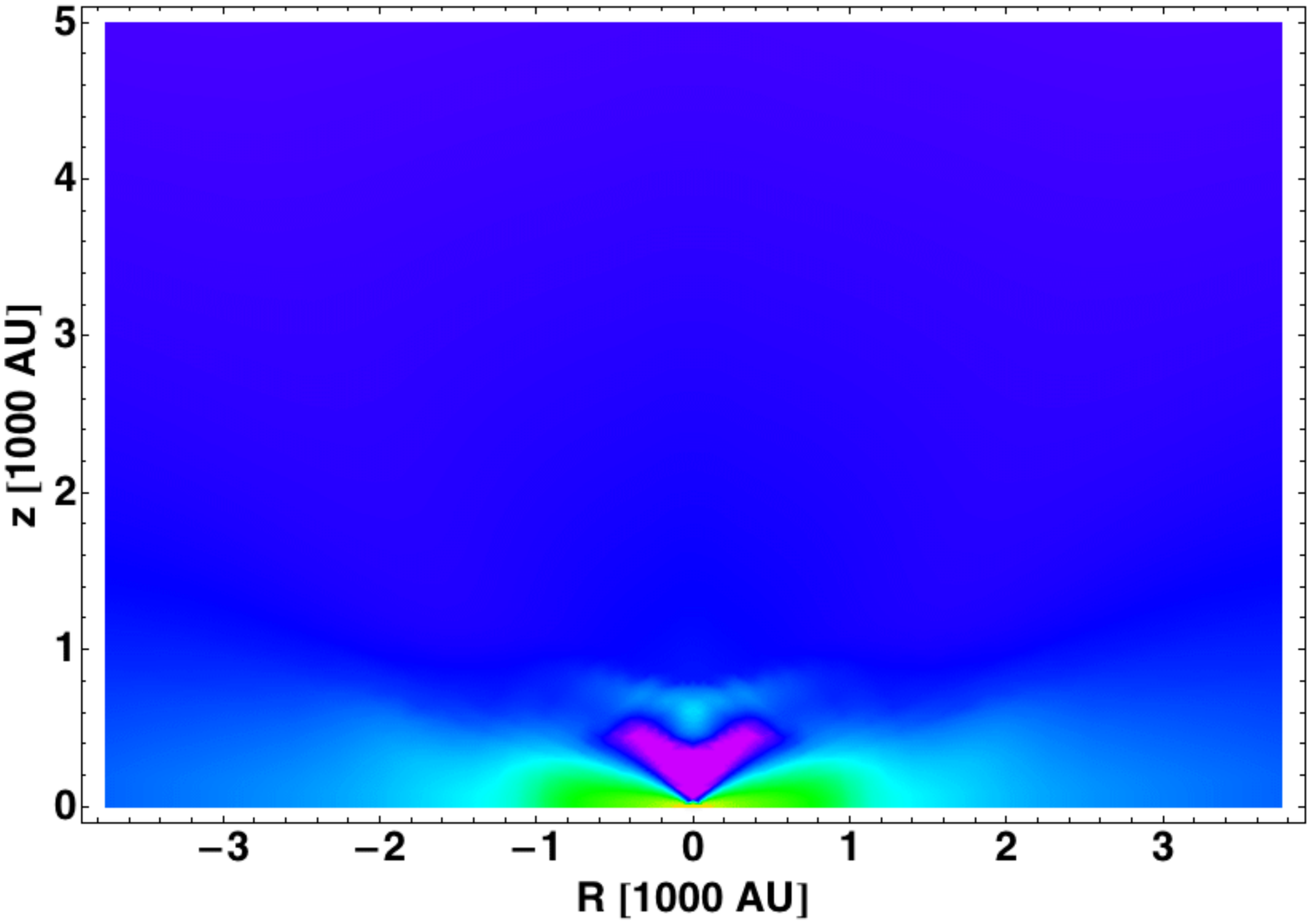}
}
\hspace{25mm}
\subfigure[RT+FLD run at $t = 39$~kyr]{
\includegraphics[width=0.36\textwidth]{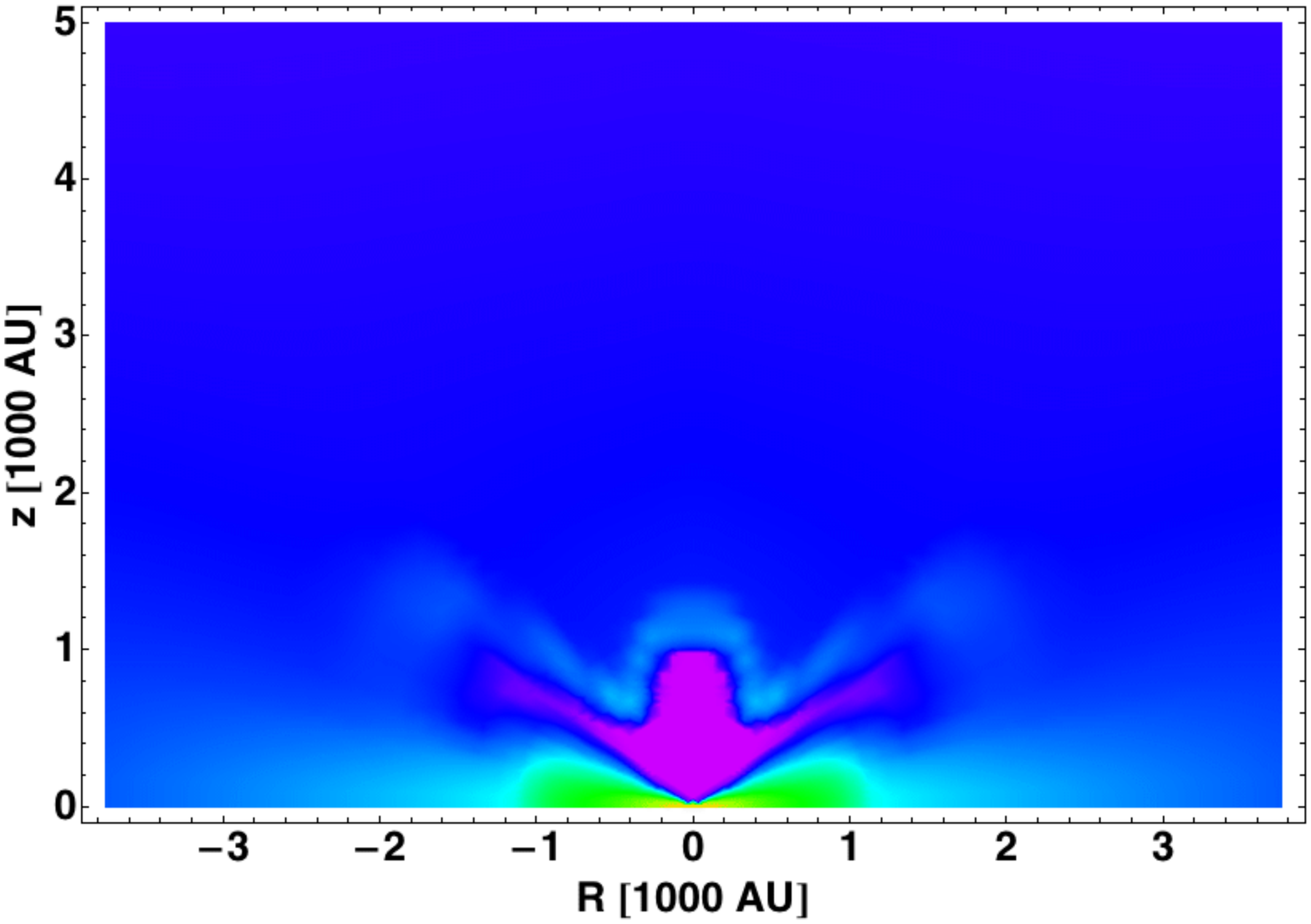}
}\\
\subfigure[FLD run at $t = 49$~kyr]{
\includegraphics[width=0.36\textwidth]{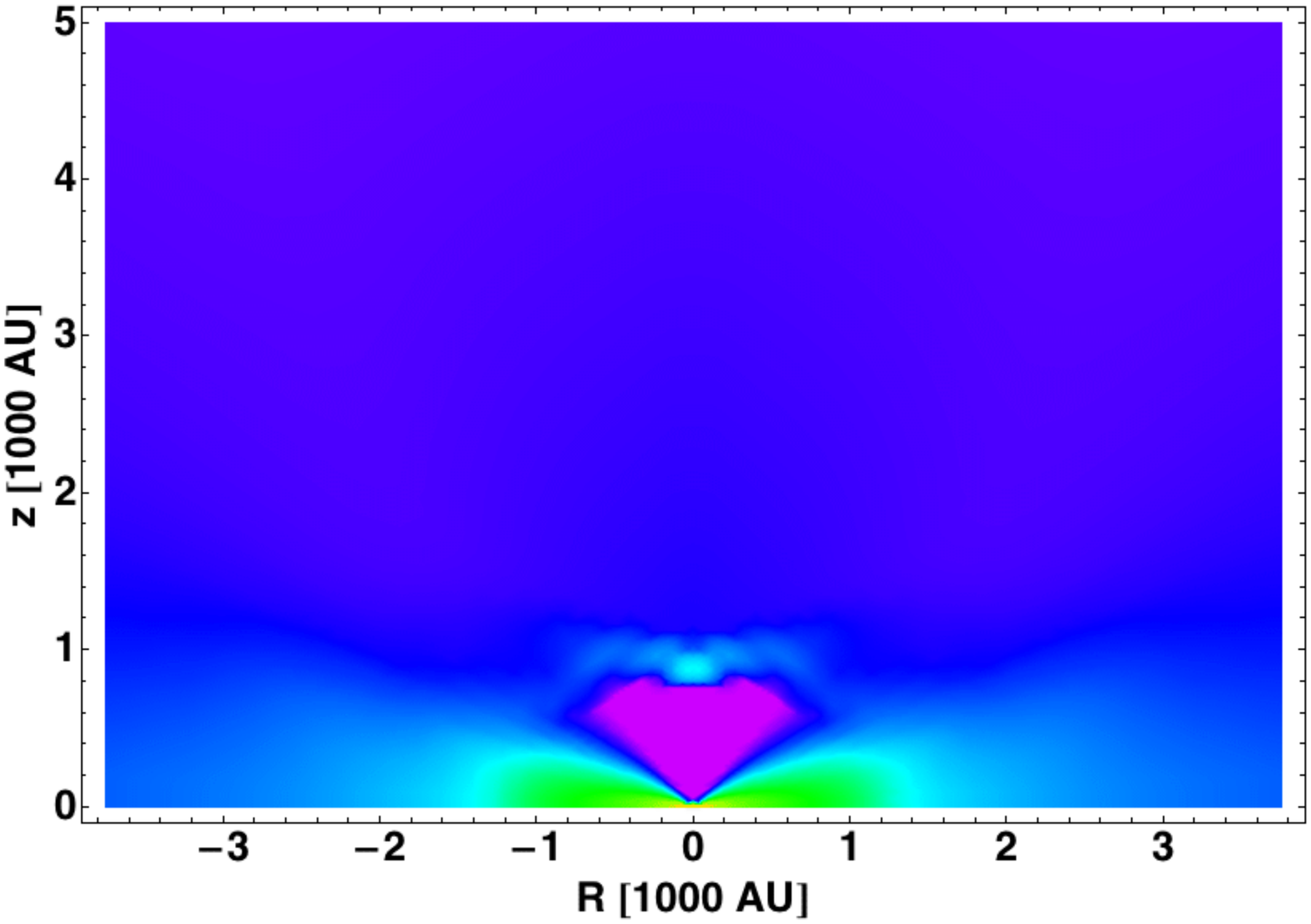}
}
\hspace{25mm}
\subfigure[RT+FLD run at $t = 44$~kyr]{
\includegraphics[width=0.36\textwidth]{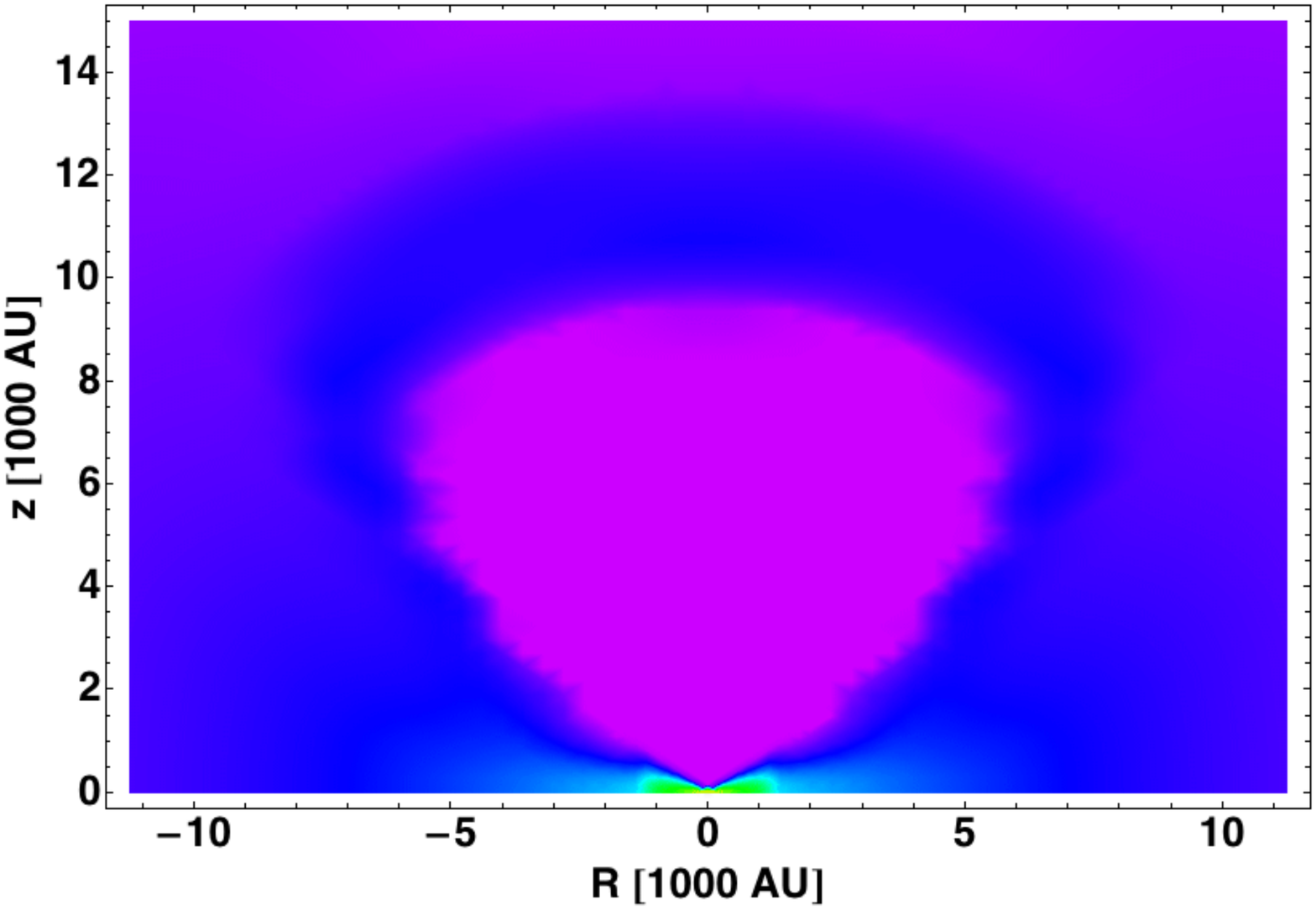}
}\\
\subfigure[FLD run at $t = 54$~kyr]{
\includegraphics[width=0.36\textwidth]{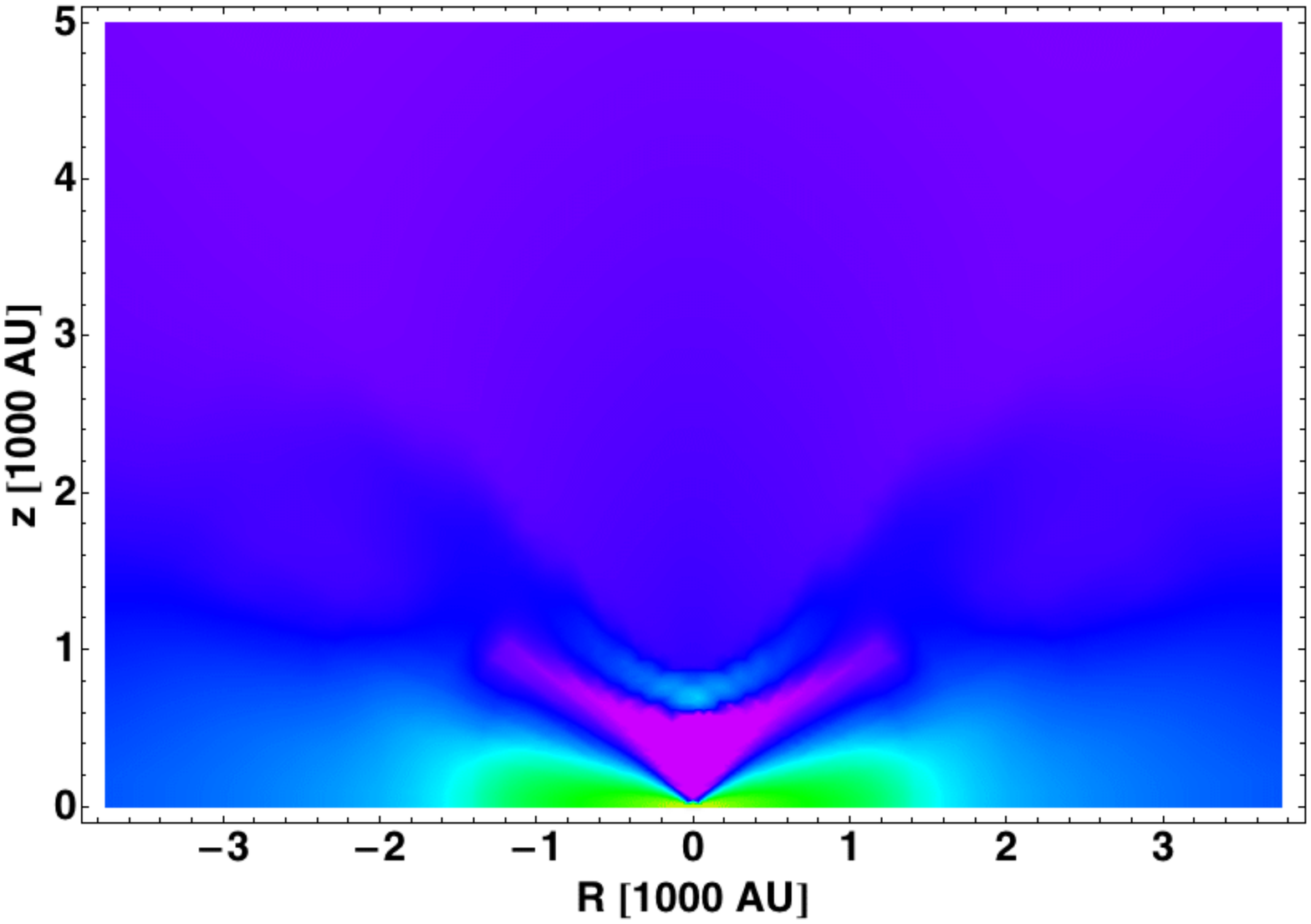}
}
\hspace{25mm}
\subfigure[RT+FLD run at $t = 49$~kyr]{
\includegraphics[width=0.36\textwidth]{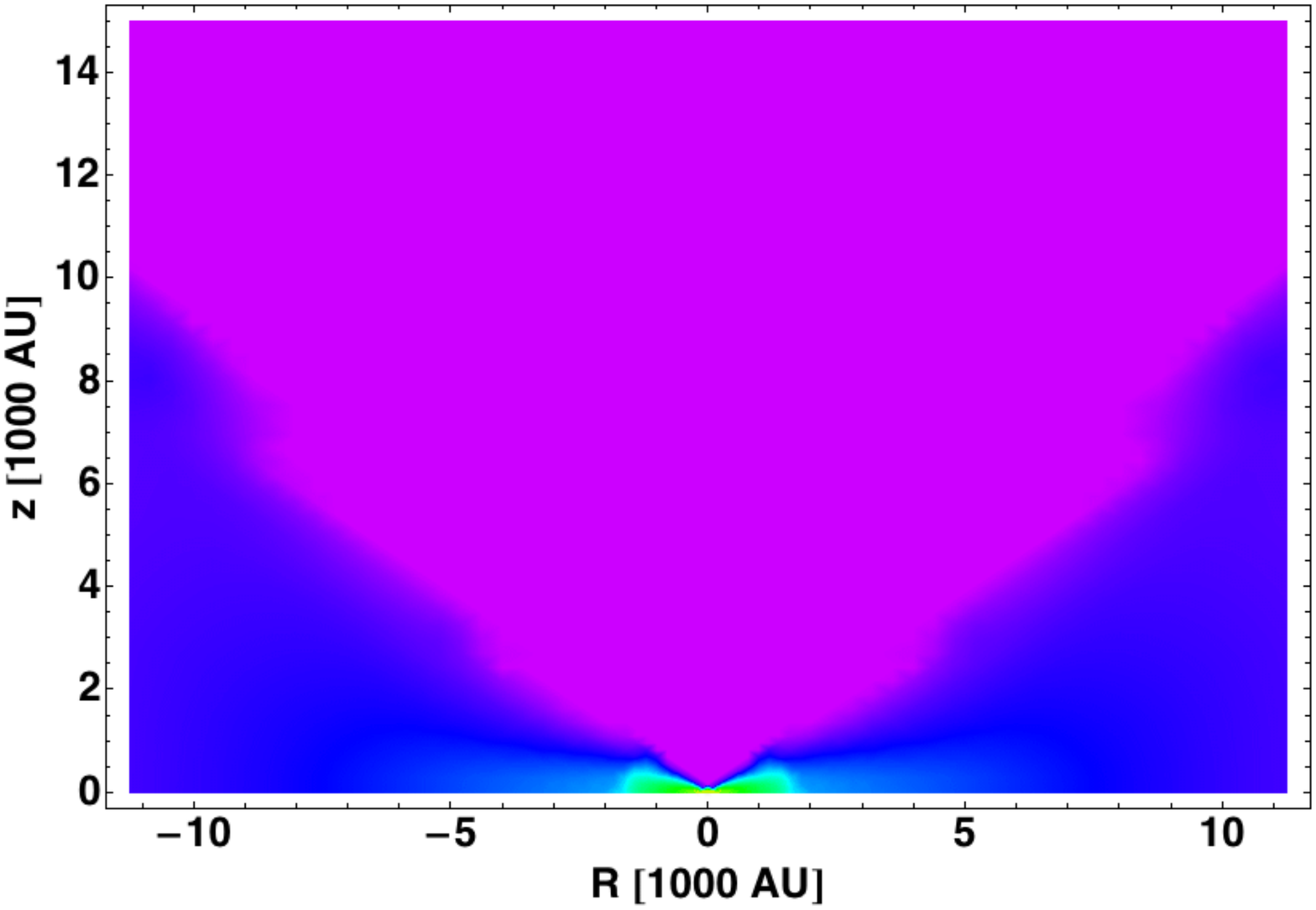}
}\\
\caption{
Simulation snapshots of the gas density for one of the FLD (left panels) and one of the ray-tracing + FLD (right panels) runs for three different points in time during the launch of the radiation pressure dominated cavities.
The data is taken from the runs with initial core mass $M_\mathrm{core} = 50\Msol$ and a radial density slope of $\beta=-2$.
\newline
Note: The spatial section of the RT+FLD snapshots increases with time to follow the rapid expansion of the outflow cavity.
\newline
Animations of the launch of the radiation-pressure-dominated cavities in the simulations are available as online material, too.
}
\label{fig:Snapshots}
\end{figure*}

In addition to the different growth rates, the radiation-pressure-driven outflow cavities also develop pronounced differences in their morphology depending on the applied radiation transport method.
As an example, the different morphologies produced by the different radiation transport methods are visualized in three snapshots in time during the onset of the outflow launching in Fig.~\ref{fig:Snapshots}.
In the case of FLD, the outflow is easily stopped by the infalling matter along the polar axis, while in the RT+FLD case the outflow cavity 
\vONE{grows rapidly with}
time.
In the FLD approximation, the radiative flux tends to follow a path that minimizes the optical depth and hence avoids the swept-up mass on top of the cavity.
This avoidance is alleviated by the centrifugal forces, which diminish the gravitational attraction in regions slightly above the disk.
In the ray-tracing method, the isotropic stellar irradiation flux directly impinges onto the swept-up mass on top of the polar cavity and pushes the mass to larger radii.
The resulting large-scale morphology of the cavity is far more isotropic. 
The opening angle of the cavity is determined by the inner disk structure.

In the following Sects.~\ref{sect:QuantitativeResults1} and \ref{sect:QuantitativeResults2}, we analyze quantitatively the outcome of the simulations presented here.
We check our hypothesis of epochs of marginal Eddington equilibrium in the FLD-only runs and determine via analytical estimates (Sect.~\ref{sect:Analytic}), why the cavity shells in the RT+FLD case do not undergo these epochs and therefore remain stable with respect to the radiative Rayleigh-Taylor instability.

\section{Quantitative analysis of the cavity growth}
\label{sect:QuantitativeResults1}
\begin{figure*}[htbp]
\begin{center}
\includegraphics[width=0.33\textwidth]{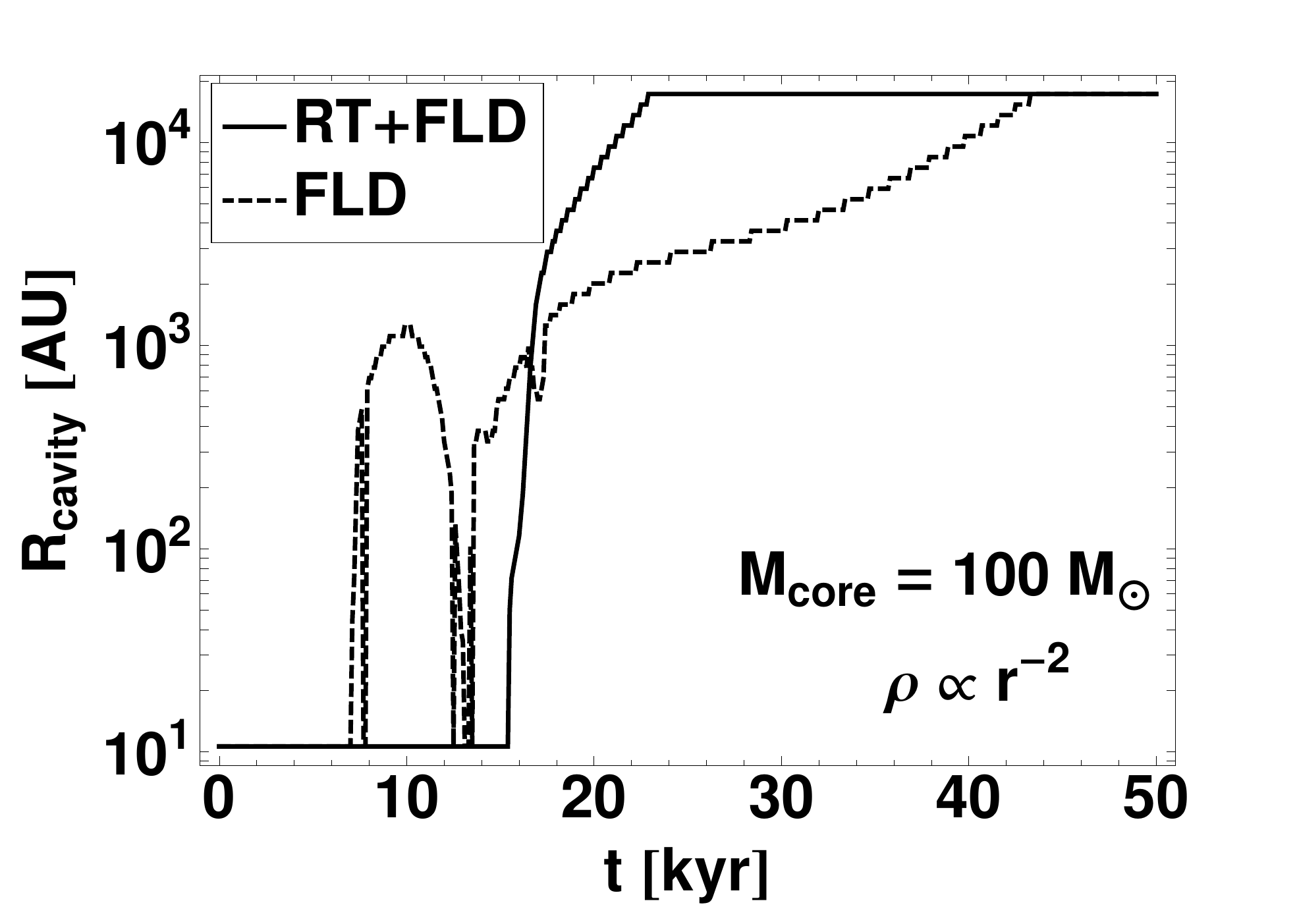}
\includegraphics[width=0.33\textwidth]{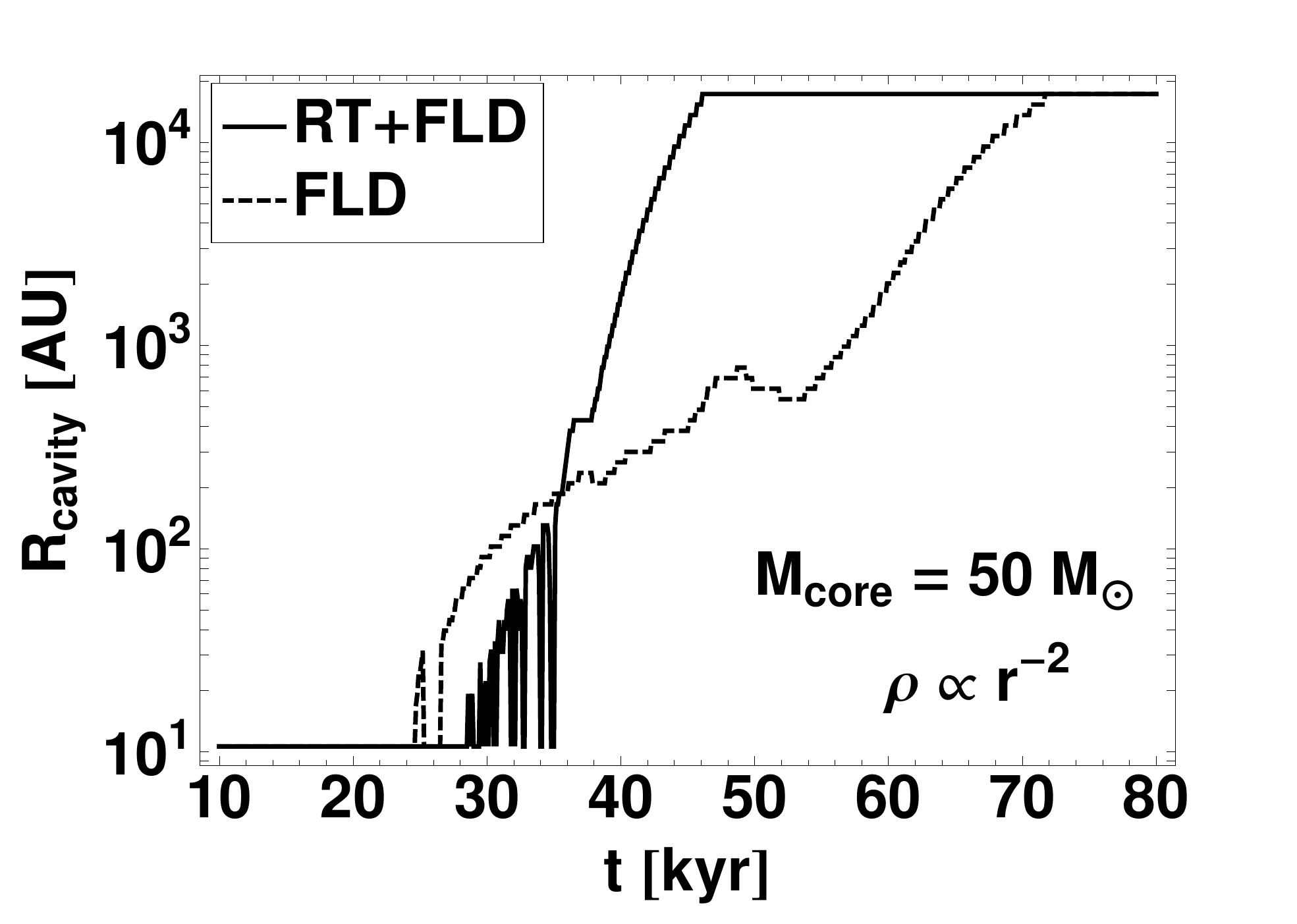}
\includegraphics[width=0.33\textwidth]{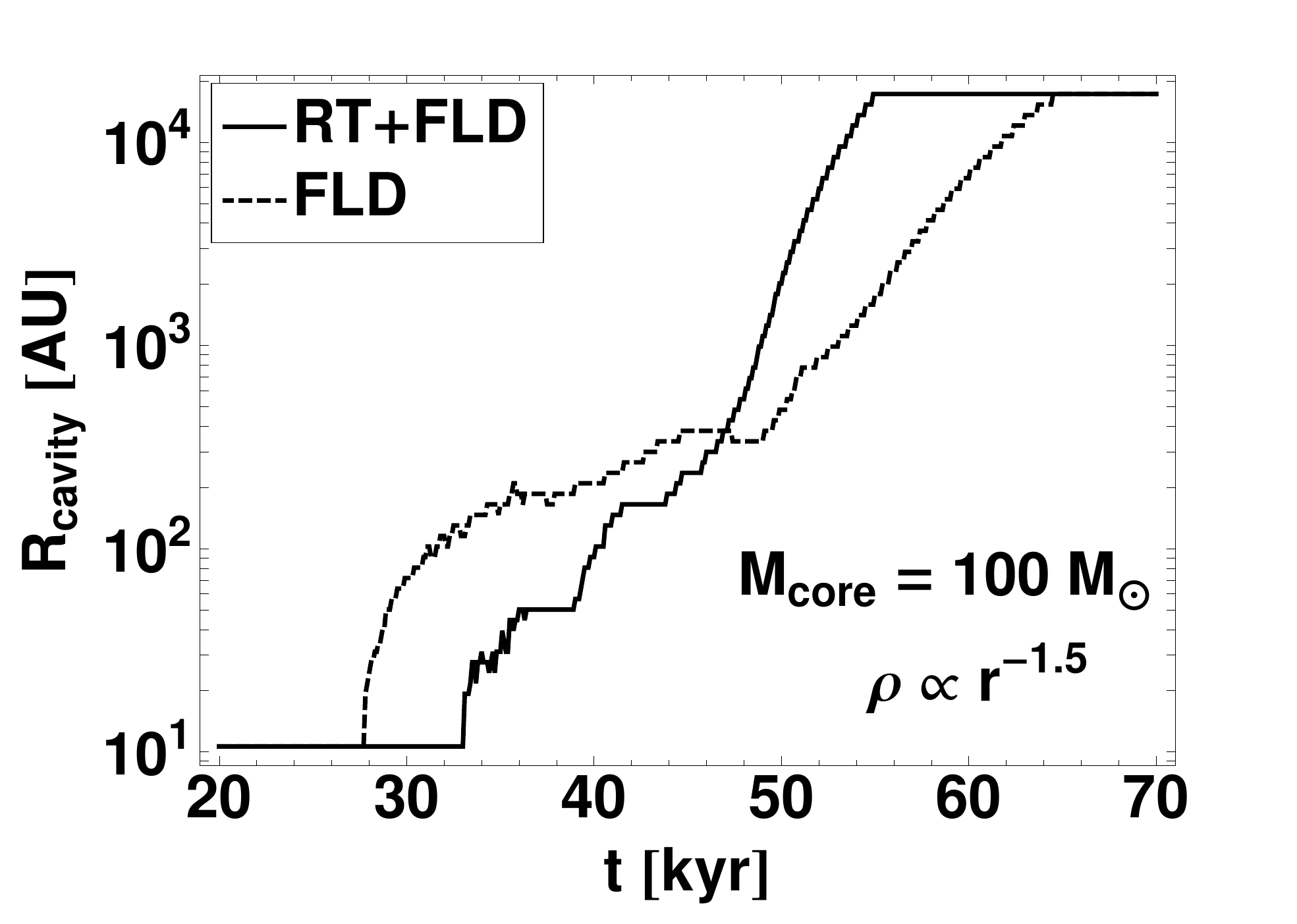}
\caption{
Size of the cavity $R_\mathrm{cavity}$ as a function of time $t$ for the three different initial conditions.
All runs were performed using the RT+FLD hybrid radiation transport scheme (solid lines) and the FLD scheme (dashed lines).
The lowest and highest extent of $R_\mathrm{cavity}$ of 10~AU and 0.1~pc, respectively, are given by the inner and outer rim of the computational domain.
}
\label{fig:OutflowRadiusvsTime}
\end{center}
\end{figure*}

We analyze quantitatively the time-dependent extent of the outflow cavity.
The radial extent $R_\mathrm{cavity}$ of the outflow cavities above the central star is determined as the extent of the cleared cavity along the polar axis, as visualized in Fig.~\ref{fig:Snapshots}.
The resulting cavity radii as a function of time are shown for the three different initial conditions as well as the different radiation transport methods in Fig.~\ref{fig:OutflowRadiusvsTime}.

For all initial conditions, the extent of the radiation pressure dominated cavities increases far more rapidly, if the stellar source of radiation is treated via a ray-tracing scheme (labeled ``RT+FLD'') than simply included in the flux-limited diffusion solver (labeled ``FLD'').
If using the FLD approximation, the outflow is launched a little bit earlier in time and after its initial growth phase undergoes an epoch of marginal stability, in which the radiation pressure force along the polar axis seems to be balanced by gravity;
the cavity growth along the polar axis is first stopped and then reversed.
Under these conditions of marginal Eddington equilibrium $\left(f_\mathrm{rad} \lesssim f_\mathrm{grav}\right)$ with FLD, the cavity shell region has been proposed to be subject to the radiative Rayleigh-Taylor instability \citep{Jacquet:2011p18452}.
As a consequence, subsequent growth phases of the outflows in the FLD runs depend on the details of the subgrid models;
the dynamics of the marginally balanced cavity shells are strongly influenced by the stellar evolution model and the dust model, particularly the treatment of sublimation and evaporation.

In contrast to this epoch of marginal Eddington equilibrium, the extent of the outflow cavity in the RT + FLD simulations increases rapidly in time.
With the exception of the $M_\mathrm{core}=50\Msol$ case, in which a correspondingly lower mass star (yielding much lower luminosity) is formed, the outflow cavity increases in size monotonically in the RT + FLD simulations.

The difference in the growth rate for the various radiation transport schemes is most prominent in the $M_\mathrm{core}=100\Msol$ and $\rho \propto r^{-2}$ case (Fig.~\ref{fig:OutflowRadiusvsTime} left panel), i.e. in the case of the initially highest mass in the central core region, allowing for a rapid formation and growth of the central star.
In the $M_\mathrm{core}=50\Msol$ case (Fig.~\ref{fig:OutflowRadiusvsTime} middle panel), the much lower stellar luminosity results in an initially unstable cavity region even in the RT+FLD case (but with a smaller extent than 100~AU).
In the $\rho \propto r^{-1.5}$ case (Fig.~\ref{fig:OutflowRadiusvsTime} right panel), the mass is initially more concentrated at larger radii and hence the cavity shell has to sweep up far more mass during its increase.
Therefore, the growth phase of the cavity takes much longer than in the cases of initially steeper density profiles.

\section{Quantitative analysis of the cavity shell morphology and dynamics}
\label{sect:QuantitativeResults2}
To unveil the physical background of the qualitative and quantitative difference of the outflow cavity  structure in the simulations using either the RT+FLD or the FLD method, we now investigate the morphology and dynamics of the cavity shell depending on the radiation transport method.
In this section, we analyze quantitatively the difference in the morphology and the dynamics of the cavity shell in the RT+FLD and FLD cases.
We focus on the data of the simulation with an initial core mass $M_\mathrm{core} = 100\Msol$ and a radial density slope of $\beta=-2$.

\begin{figure*}[htbp]
\begin{center}
\includegraphics[width=0.49\textwidth]{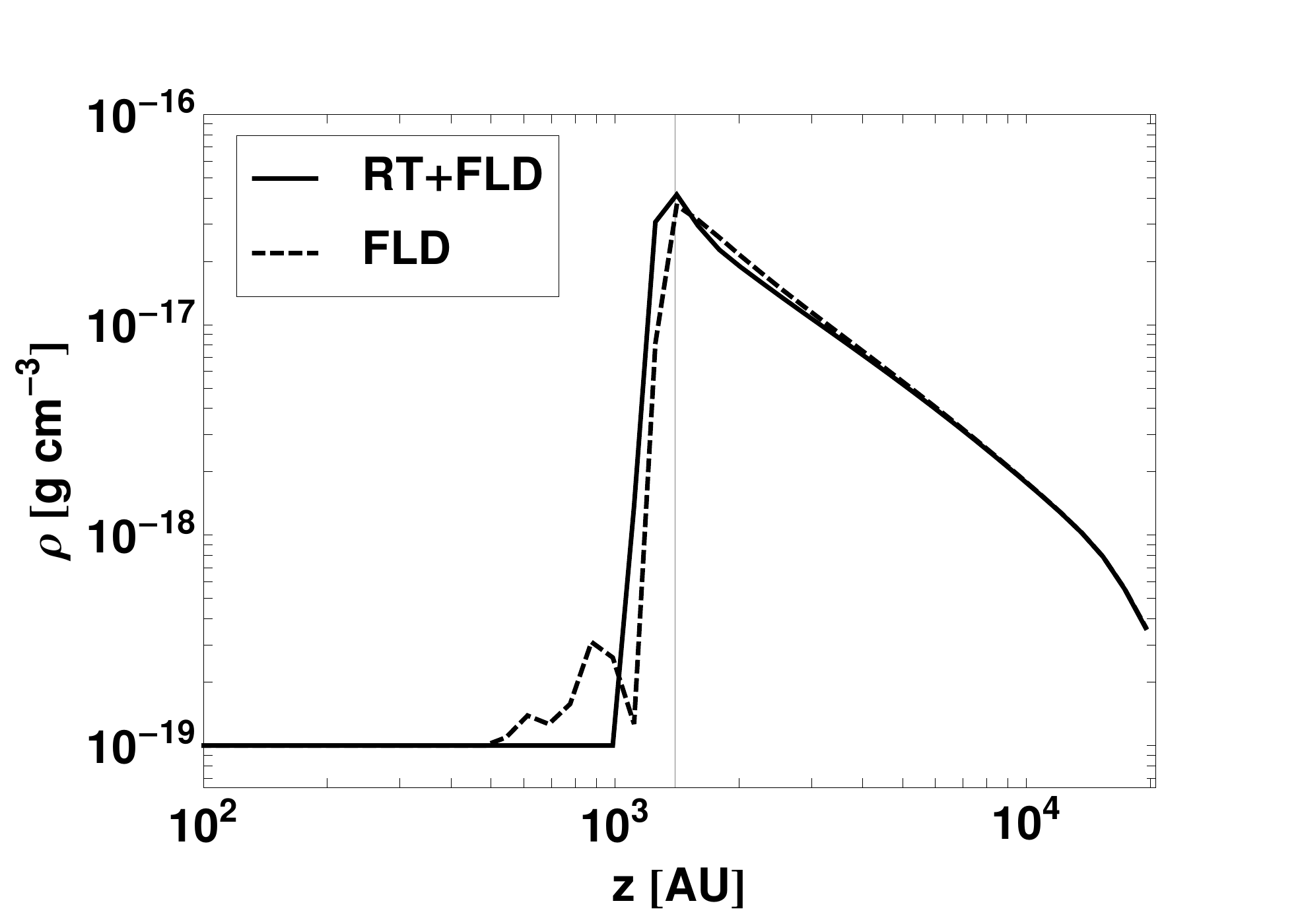}
\includegraphics[width=0.49\textwidth]{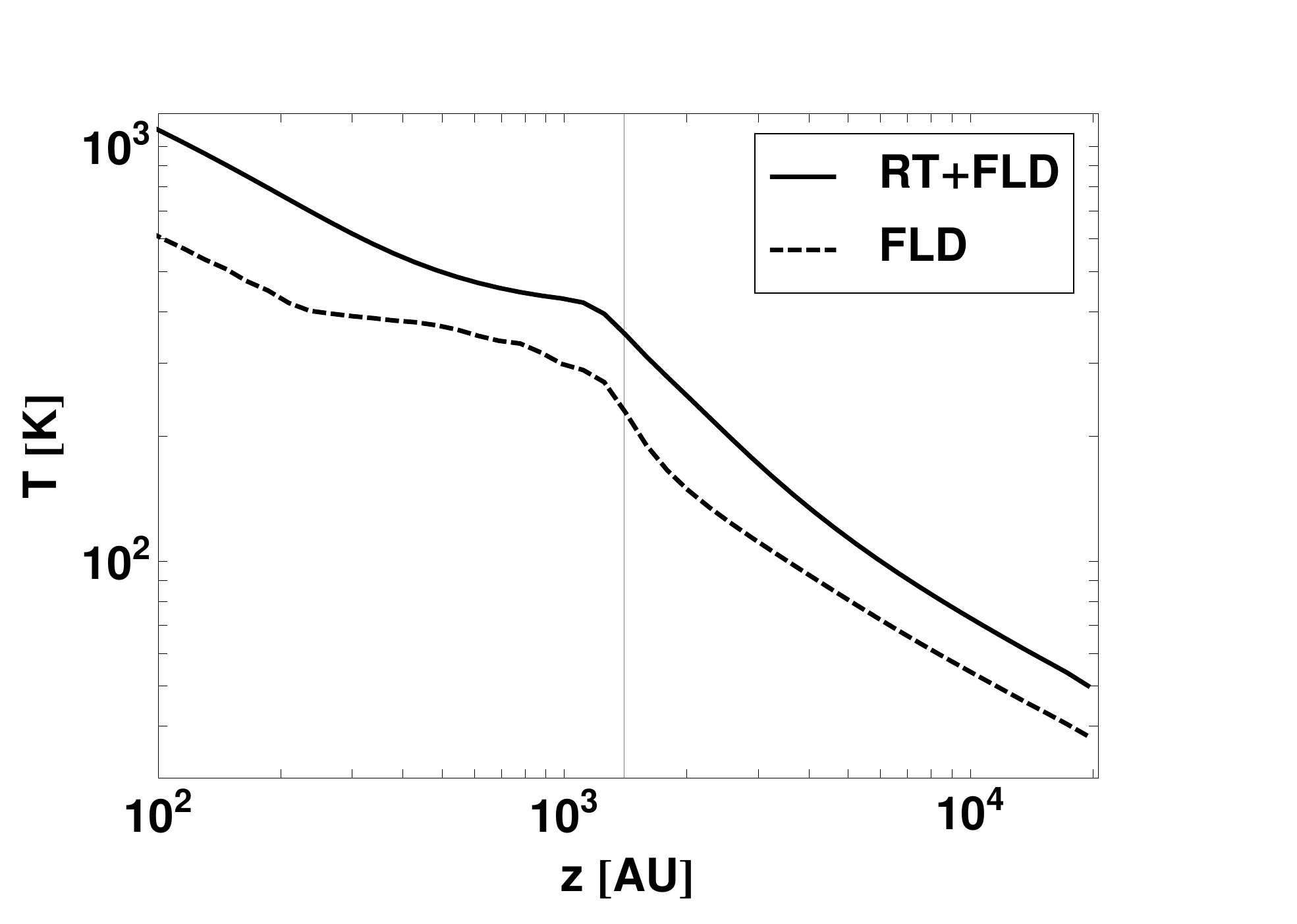}\\
\includegraphics[width=0.49\textwidth]{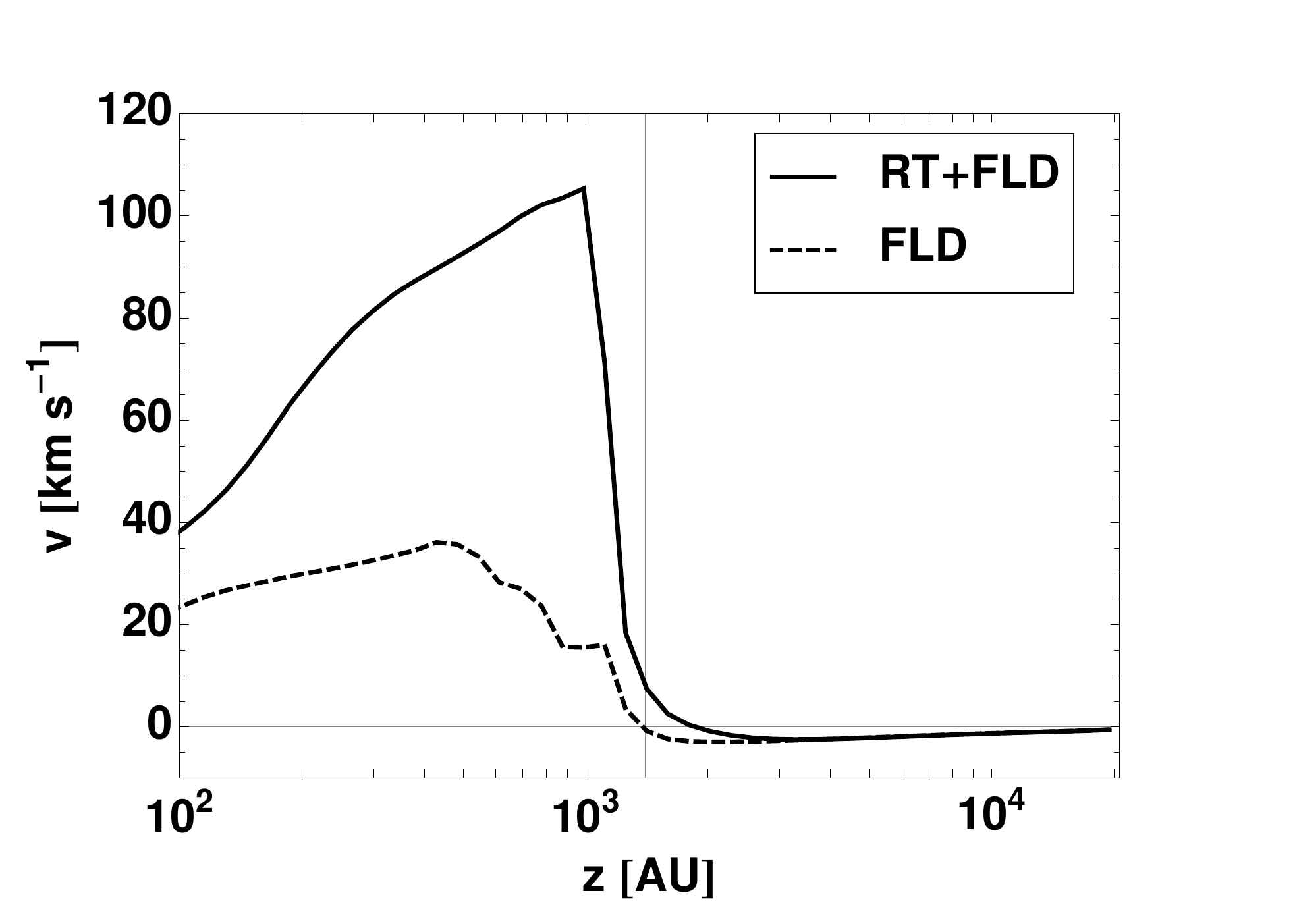}
\includegraphics[width=0.49\textwidth]{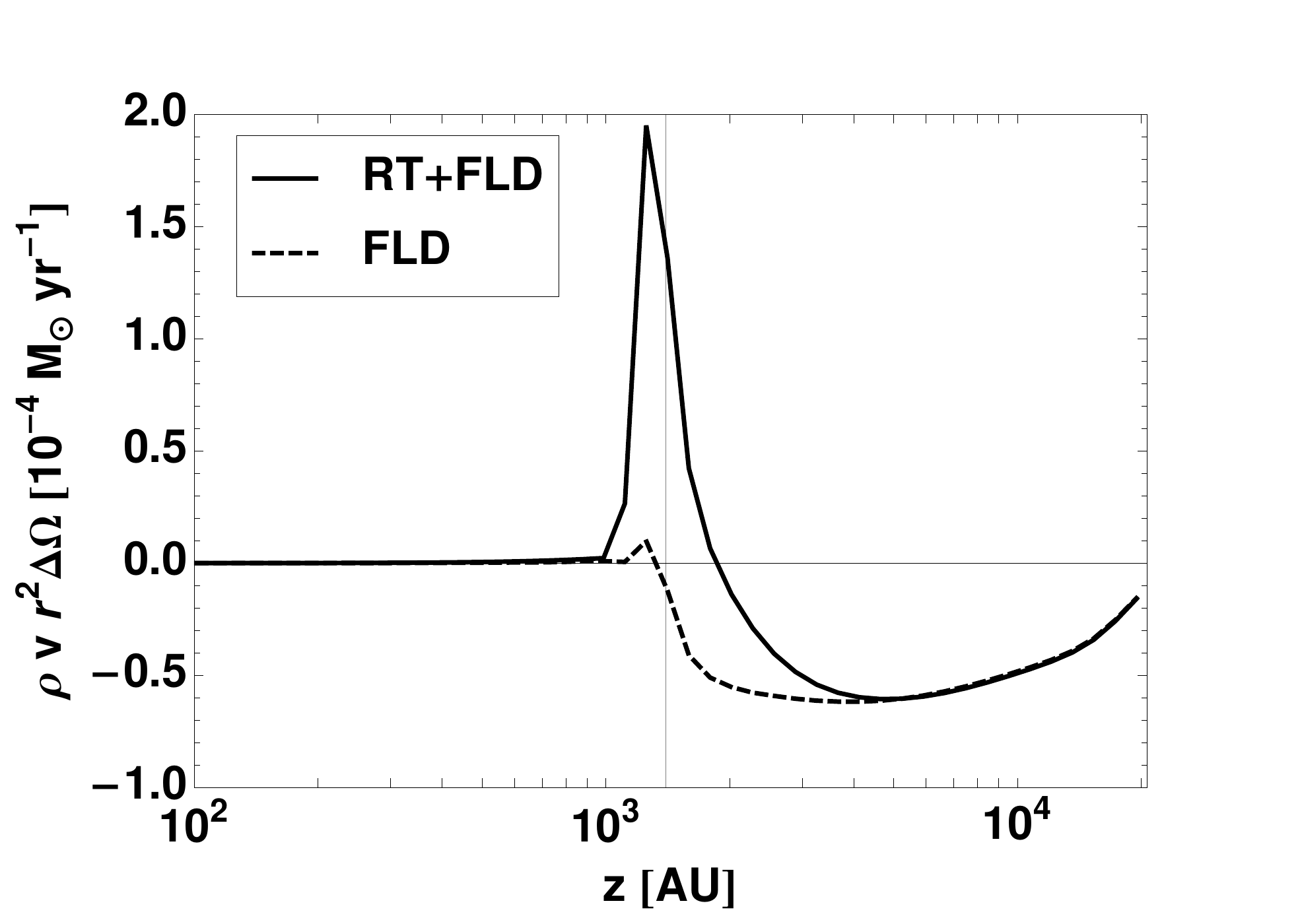}
\caption{
Gas density $\rho$ (upper left panel), 
temperature $T$ (upper right panel),  
radial velocity $v$ (lower left panel), and
radial mass-loss rate (lower right panel) 
as a function of height $z$ above the star along the polar axis.
Data is for the run with initial core mass $M_\mathrm{core} = 100\Msol$ and radial density slope $\beta=-2$.
The snapshot in time ($t=16.7$~kyr) is chosen in such a way that the cavity shell has arrived at the same location in both simulations (RT+FLD and FLD only).
The lowest density of $\rho=10^{-19}\rhocgs$ in the upper left panel is given by the numerical floor value of the hydrodynamics solver.
}
\label{fig:1Danalysis}
\end{center}
\end{figure*}

The cavity in the FLD case increases in its first growth phase up to roughly 1400~AU before the expansion along the polar axis stops and the mass flow is reversed by gravity.
We analyze the gas dynamics at the point in time when in both simulations (with RT+FLD and with FLD only) the expansion of the outflow cavity has arrived at the same location, namely at $t=16.7$~kyr.
The gas density, temperature, radial velocity, and the radial mass loss rate along the polar axis at this time are shown in Fig.~\ref{fig:1Danalysis}.
The upper left panel of this figure shows that there is almost no difference in the radial extent, compactness, and peak density of the swept-up mass in the cavity shell. 
The peaks in the density distributions correspond to $4.1\times10^{-17}\rhocgs$ and $3.7\times10^{-17}\rhocgs$ in the RT+FLD and FLD cases, respectively.

The upper right panel of Fig.~\ref{fig:1Danalysis} shows that the RT+FLD run results in a continuously slightly higher temperature in the cleared cavity, the cavity shell, and beyond.
The temperature distribution in the pure FLD run stays 24\% to 45\% below the temperature distribution of the RT+FLD run.

In contrast to these similarities, the two lower panels of Fig.~\ref{fig:1Danalysis} highlight the striking difference of both runs in their radial velocities and mass flux along the polar axis.
In the RT+FLD run, the swept-up shell material moves into the interstellar medium with a speed slightly higher than 100~km~$\mbox{s}^{-1}$, which is roughly a factor of three higher than in the corresponding FLD run.
At larger radii ($r > 3000$~AU), the velocity slope is still dominated by gravitational infall and hence \vONE{is} independent of the radiation transport method.

Up to the onset of the Eddington equilibrium epoch in the FLD run, the mass of the central star in both simulations (RT+FLD and FLD only) closely match, leading to the same strength of gravitational attraction for the matter in the cavity and shell towards the direction of the star.
Furthermore, the density structure is roughly the same (Fig.~\ref{fig:1Danalysis}, upper left panel) and the deviation in temperature within the cavity is smaller than 24-45\% (Fig.~\ref{fig:1Danalysis}, upper right panel), hence the slightly higher thermal pressure cannot be the main driver of the enormous difference in the radial velocities and mass flux (Fig.~\ref{fig:1Danalysis}, lower panels).

As a consequence, the radiation pressure acting on the swept-up mass on top of the outflow cavity has to differ in the two radiation transport methods.
First, the FLD approximation assumes that the dust grains of the swept-up mass on top of the cavity are embedded in a local radiation field, whereas the RT+FLD method takes into account the (frequency-dependent) absorption probability caused by direct stellar irradiation.
The resulting difference in the radiative acceleration at the top of the outflow cavity is analytically estimated in the following section.

Secondly, since the FLD approximation is mathematically a moment method, the derivation of the FLD approximation includes the integral over the angular distribution of the radiative flux; hence the emitted photons of the central star do not move along straight rays until they are absorbed.
The radiative flux in the FLD approximation instead follows a path that minimizes the optical depth.
The FLD method was originally introduced to describe spherically symmetric flows, hence the integral over the propagation direction did not result in unphysical behavior \citep[see e.g.][]{Bruenn:1985p18798}.
In a multi-dimensional environment with a non-isotropic optical depth, this assumption of the FLD approximation breaks down. 
Including higher-order terms in the derivation would minimize this inaccuracy and potentially yield a sufficient tracing of the correct photon path.
However, in the RT+FLD scheme the irradiation by the central star is computed via a tay-tracing equation, which takes into account the correct photon propagation direction, i.e.~the stellar irradiation flux directly impinges on the swept-up mass shell at the top of the optically thin cavity.
\begin{figure*}[t]
\begin{center}
\includegraphics[width=0.49\textwidth]{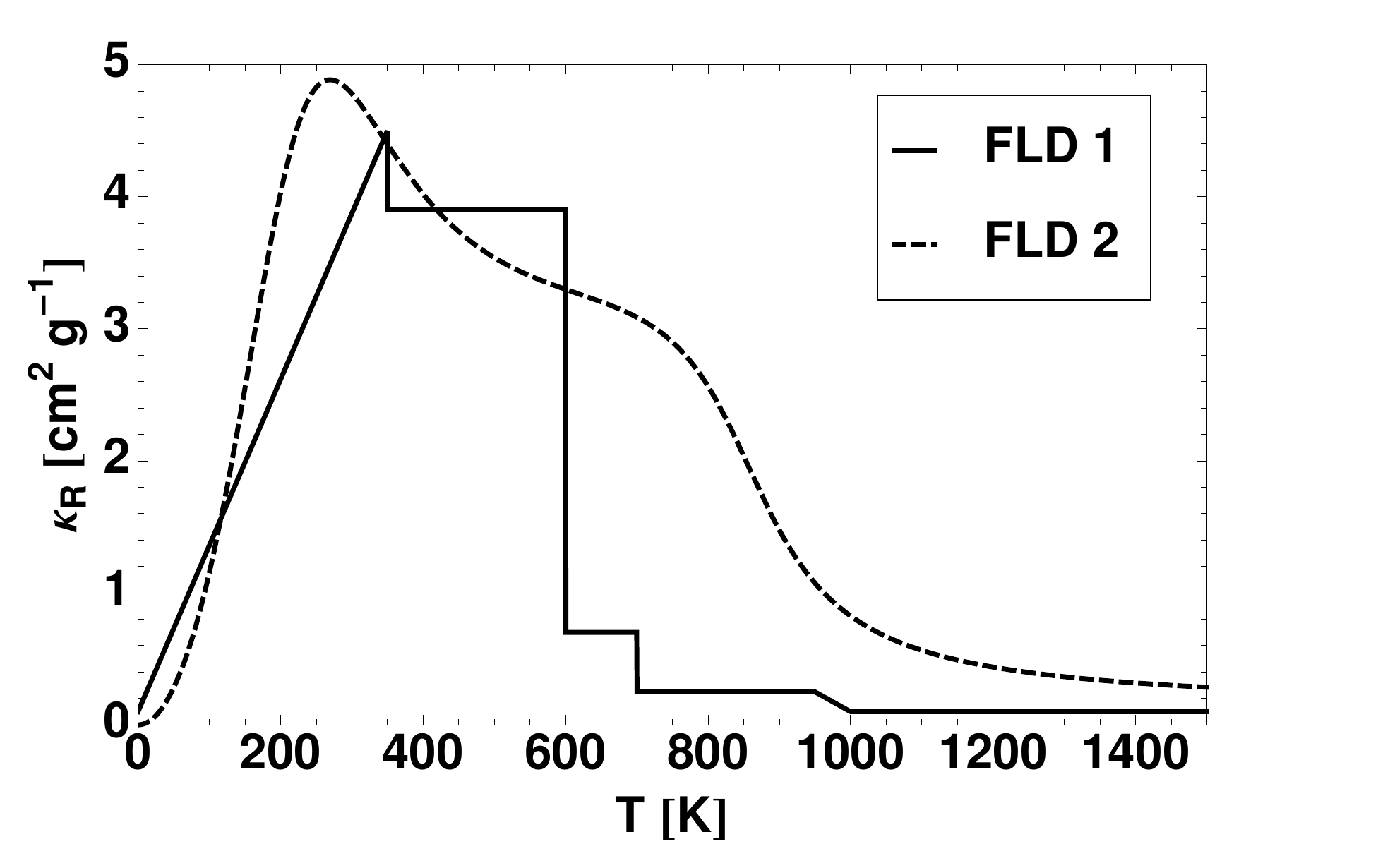}
\includegraphics[width=0.49\textwidth]{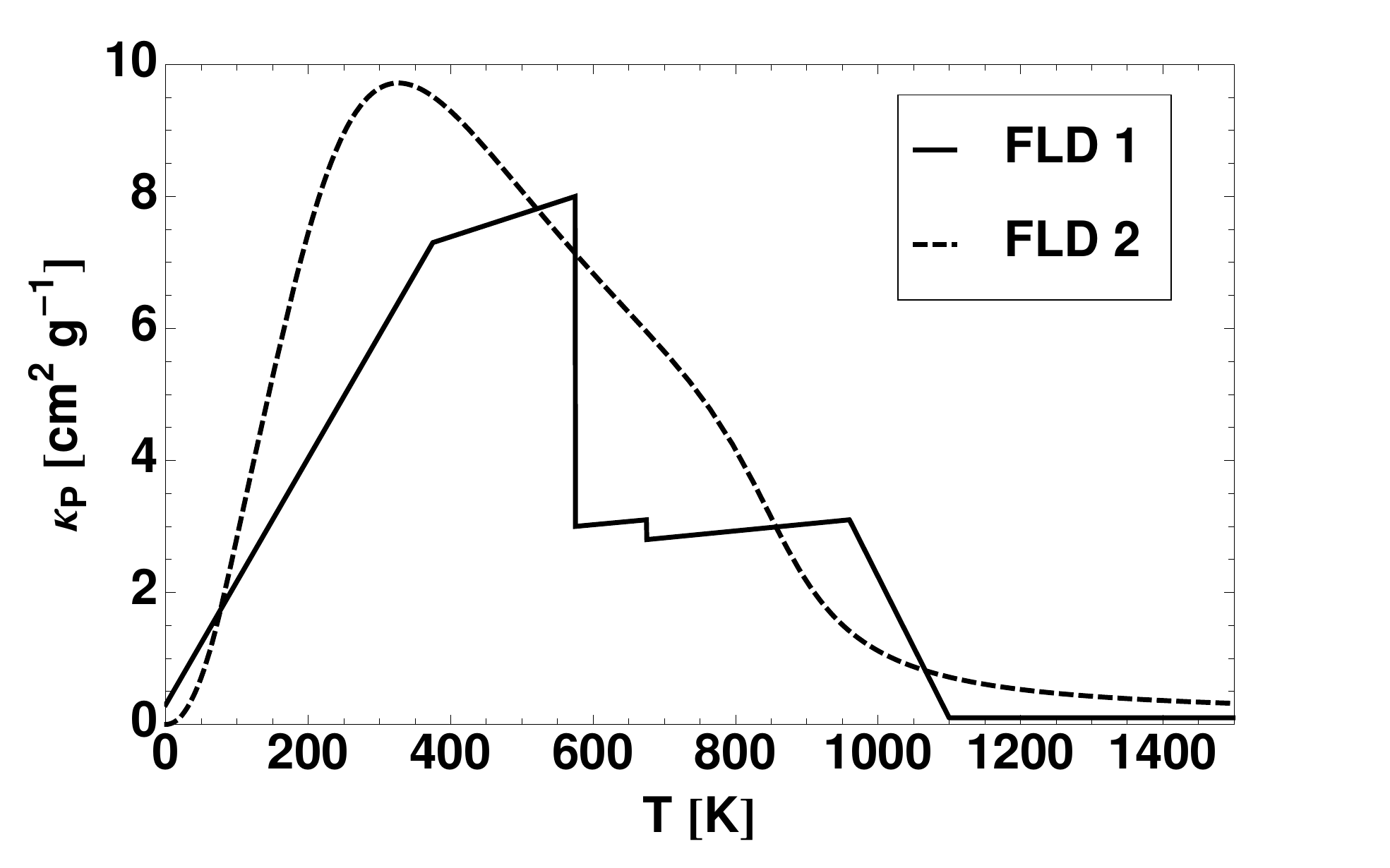}
\caption{
Rosseland mean opacities $\kappa_\mathrm{R}$ (left panel) and 
Planck mean opacities $\kappa_\mathrm{P}$ (right panel) 
per gram gas as a function of the local radiation temperature $T$.
The label ``FLD 1'' denotes the opacity description by \citet{Krumholz:2007p1380}, Eq.~(11) therein.
The label ``FLD 2'' denotes the opacities used in \citet{Kuiper:2010p17191, Kuiper:2011p17433} as well as in this paper;
these opacities are computed based on the density-dependent evaporation of dust grains using the low density $\rho=10^{-19}\rhocgs$ of the polar cavities.
}
\label{fig:OpacitiesFLD}
\end{center}
\end{figure*}

\section{Analytical determination of the acceleration caused by stellar irradiation in the ray-tracing approach and the flux-limited diffusion approximation}
\label{sect:Analytic}
Our aforementioned analysis indicates that the difference between the cavity shell stability in the two radiation transport methods is due to the higher radiative acceleration of the swept-up material when the RT approach is used for the stellar irradiation instead of the FLD approximation.
The use of RT guarantees that the stellar radiation directly interacts with the swept-up mass at the top of the optically thin outflow cavity.
The dust grains in the swept-up material therefore absorb the stellar flux following the frequency-dependent absorption coefficient $\kappa(\nu)$.
If we integrate over frequencies, the dust grains absorb the stellar spectrum with respect to the Planck mean opacity $\kappa_\mathrm{P}(T_*)$ at a temperature $T_*$ of the stellar photosphere.
\vONE{In contrast,}
the FLD approximation assumes that dust grains are embedded in a locally isotropic radiation field and that their absorption behavior is based on the Rosseland mean opacity $\kappa_\mathrm{R}(T)$ at the local radiation temperature $T$ at the top of the cavity.
Computing the Planck or Rosseland mean opacity for a given temperature $T$ mostly results in a difference of only a factor of two. 
However, computing the Planck mean opacity with the stellar temperature $T_*$ instead of the local radiation temperature $T$ leads to enormous differences.
In the next subsection, we determine and illustrate the difference between these opacities.

\subsection{Opacities}
For the FLD part of our hybrid radiation transport scheme, we compute the Rosseland mean opacities $\kappa_\mathrm{R}(T)$ directly from the frequency-dependent opacity table \citep{Laor:1993p736}.
To obtain the absorption coefficient in a given grid cell, these opacities are multiplied by the local dust-to-gas mass ratio $M_\mathrm{dust} / M_\mathrm{gas}(\vec{x})$.
In our models, the evaporation temperature of dust grains is calculated depending on the local gas density by applying the formula of \citet{Isella:2005p3014}, which represents a power-law fit to the \citet{Pollack:1994p3016} data.
For the opacities in the simulation of \citet{Krumholz:2009p10975}, the authors use linear regressions to the \citet{Pollack:1994p3016} data in a small couple of intervals;
the density dependence of the evaporation of dust grains is not taken into account.

The Rosseland $\kappa_\mathrm{R}(T)$ and Planck $\kappa_\mathrm{P}(T)$ mean opacities used in our simulations and those applied in \citet{Krumholz:2007p1380}, Eq.~(11), are visualized in Fig.~\ref{fig:OpacitiesFLD}.
Owing to the similarities \vONE{between} our and the Krumholz et al.~opacities,
the resulting radiation pressure in the cavity and shell should be comparable in the runs using the FLD radiation transport method.
If there were a difference at all, the higher mean opacities in our simulations for $T>600$~K would lead to a slightly higher radiation pressure onto the swept-up mass shell than in the simulations of Krumholz et al.
If our opacities in the FLD run lead to an unstable cavity shell, the opacities used by Krumholz et al.~have to yield the same outcome.

\begin{figure}[htbp]
\begin{center}
\includegraphics[width=0.49\textwidth]{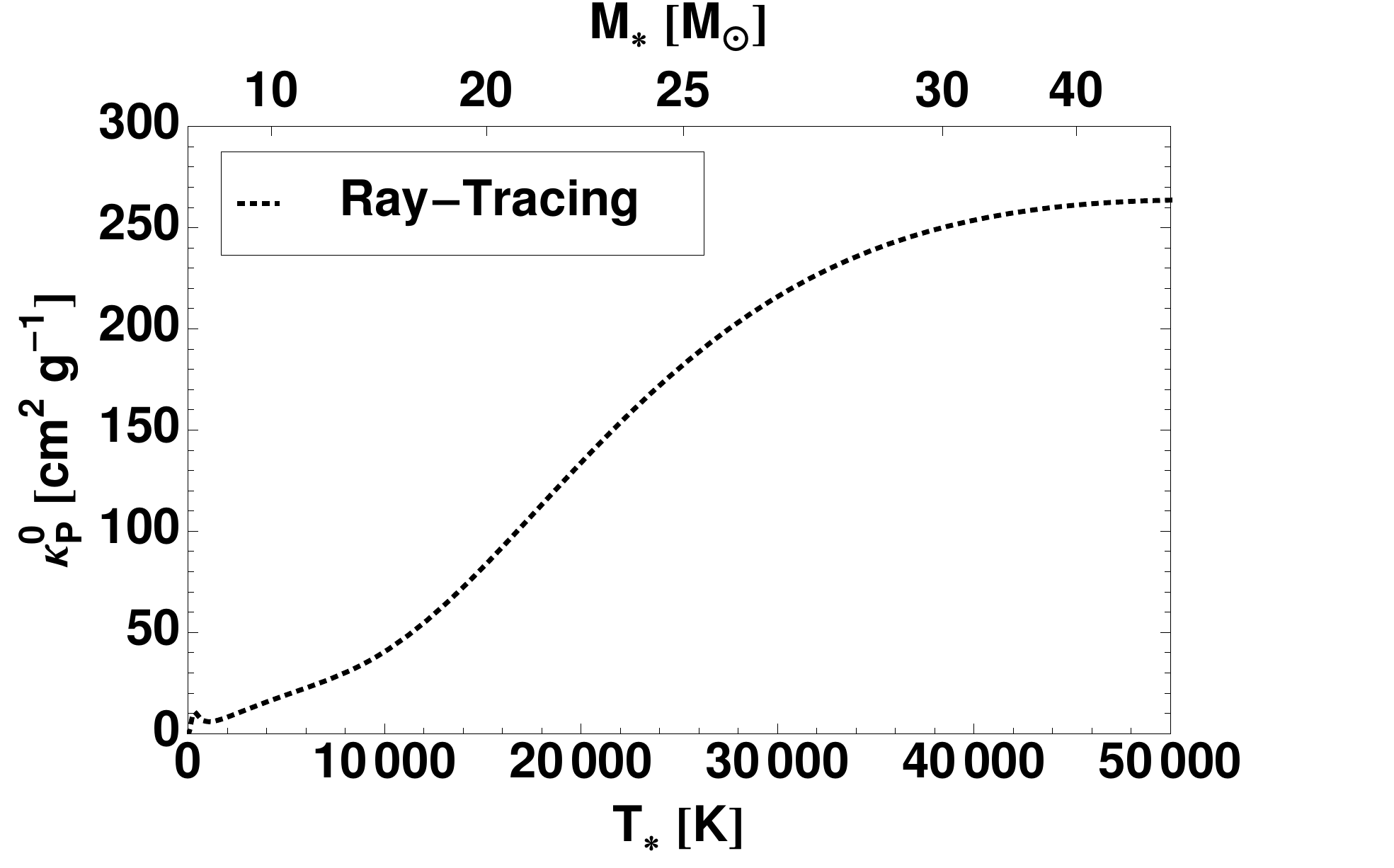}
\caption{
Planck mean opacities $\kappa_\mathrm{P}^0$ per gram gas as a function of the radiation temperature $T_*$ of the stellar photosphere
as used in the RT step of the hybrid radiation transport scheme in \citet{Kuiper:2010p17191, Kuiper:2011p17433} and in this paper.
The upper horizontal axis shows the mass of the proto-star, which corresponds to the stellar surface temperature $T_*$, taken from the \citet{Hosokawa:2009p12591} evolutionary track for a constant accretion rate of $\dot{M}=10^{-3}\Msol~\mathrm{yr}^{-1}$.
}
\label{fig:OpacitiesRT}
\end{center}
\end{figure}

In our RT+FLD runs, the FLD approximation is only used to determine the thermal radiation by dust grains.
In addition, the irradiation of the stellar luminosity up to the first absorption event (i.e.~up to an optical depth of $\tau(\nu) \approx 1$) is computed by a RT step that appropriately resembles the propagation and absorption of the stellar photons.
The dust grains absorb the star light according to the frequency-dependent absorption coefficient $\kappa(\nu)$.
Although this frequency dependence is fully covered in our RT solver (we use 79 frequency bins in the case of the \citet{Laor:1993p736} opacities), in this derivation -- for simplicity -- we use the gray approximation by computing the corresponding Planck mean opacities $\kappa_\mathrm{P}(T_*)$ that depend on the stellar surface temperature $T_*$.
Fig.~\ref{fig:OpacitiesRT} shows the resulting Planck mean opacities $\kappa_\mathrm{P}^0(T_*)$ as a function of the stellar surface temperature $T_*$.
To obtain the absorption coefficient $\kappa_\mathrm{P}(T_*)$ 
in a given grid cell of the computational domain for these stellar photons, the opacities $\kappa_\mathrm{P}^0(T_*)$ are multiplied by the 
evaporation probability, and, therefore, \vONE{also} depend on the local gas density and temperature.

The mass of the proto-star during the onset of the cavity shell instability in the FLD run is about $20 \mbox{ M}_\odot$ and higher.
Comparing the absorption coefficient of the dust grains in the shell on top of the optically thin cavity in the FLD approximation (Fig.~\ref{fig:OpacitiesFLD}) with the RT method (Fig.~\ref{fig:OpacitiesRT}), immediately illustrates that the FLD approximation underestimates the absorption coefficient tremendously (at least by a factor of 20-50).

In the following subsections, we show that this underestimated absorption coefficient for direct stellar irradiation in the FLD method leads to epochs of marginal Eddington equilibrium in the cavity shell, allowing the shell to become unstable, whereas the higher-order treatment of stellar irradiation via RT results in a stable cavity shell with radial velocities up to $v \gtrsim100 \mathrm{~km~s}^{-1}$ and mass-loss rates of $\dot{M} \approx 2 \times 10^{-4} \Msol \mbox{ yr}^{-1}$.

\subsection{Radiative acceleration}
To compute the radiative acceleration $\vec{a}_\mathrm{rad}$ of the swept-up material at the top of the optically thin cavity, we derive analytic expressions for the different radiation transport approaches based on formulae similar to the well-known Eddington limit.
We start with the formula for radiative acceleration 
\begin{equation}
\label{eq:Mihalas}
\vec{a}_\mathrm{rad} = \kappa \frac{\vec{F}}{c}
\end{equation}
for the radiative flux $\vec{F}$ and the speed of light $c$.
Since we only wish to estimate the radiation pressure by the first absorption of the (isotropic) stellar light and the cavity is treated as optically thin up to the location $R_\mathrm{cavity}$ of the swept-up material on top of it, we compute the radiative acceleration onto dust grains only along the direction of the polar axis (1D) and express the radiative flux by 
\begin{equation}
\label{eq:RadiativeFlux}
F(R_\mathrm{cavity}) = \frac{L_*}{4\pi~R_\mathrm{cavity}^2}.
\end{equation}
Eq.~\ref{eq:RadiativeFlux} is valid either for the assumption of an optically thin cavity or -- owing to the conservation of momentum and energy -- as long as the swept-up mass in the cavity shell can be treated as spherically symmetric.
By means of the absorption of stellar photons, the dust grains heat up and then re-emit photons at longer wavelengths.
Integrating over all photons penetrating a shell at the radius $r$ and the momentum gained by the dust grains ensures that there is an excellent conservation of momentum.

Inserting Eq.~\ref{eq:RadiativeFlux} into Eq.~\ref{eq:Mihalas} results in a radiative acceleration that depends on the absorption coefficient $\kappa$, the stellar luminosity $L_*$, and the cavity height $R_\mathrm{cavity}$ above the star
\begin{equation}
\label{eq:RadiativeAcceleration}
a_\mathrm{rad} = \kappa \frac{L_*}{4\pi~c~R_\mathrm{cavity}^2}
\end{equation}
This radiative acceleration has to be compared with the gravitational attraction $a_\mathrm{grav}$ of the dust grains at the cavity shell position $R_\mathrm{cavity}$ of the central stellar mass $M_*$
\begin{equation}
\label{eq:GravitationalAcceleration}
a_\mathrm{grav} = \frac{G~M_*}{R_\mathrm{cavity}^2},
\end{equation}
where $G$ is the gravitational constant.
The gravity of the less massive circumstellar accretion disk is neglected here.
In observations and simulations, even the mass of the accretion disk plus the rotating large-scale torus further away from the cavity shell is less than or comparable to the mass of the central star; 
hence neglecting this mass reduces the stellar gravity by only a factor of two at most.
To minimize the numbers of independent variables, we derive the stellar luminosity $L_*$ from the stellar mass $M_*$ by assuming that the proto-star follows the evolutionary track of \citet{Hosokawa:2009p12591} with a constant accretion rate of $\dot{M}=10^{-3}\Msol~\mathrm{yr}^{-1}$.
The assumption of this particular evolutionary track does of course influence the exact quantitative comparison of the radiative versus the gravitational acceleration, but the qualitative outcome does not change.
The critical factor in this comparison is the strength of the absorption coefficient $\kappa$.

In the FLD approximation, the absorption coefficient $\kappa$ is given by the Rosseland mean opacity $\kappa_\mathrm{R}(T)$ at the local radiation temperature $T$ of the cavity shell.

In the following, we compute the local gas temperature $T$ at the top of the cavity by the formula of \citet{Spitzer:1978p18454} for an optically thin stellar environment with the prerequisite $R_\mathrm{cavity} \gg R_*$ that the distance $R_\mathrm{cavity}$ to the shell is much larger than the stellar radius $R_*$ 
\begin{equation}
\label{eq:Spitzer}
T(R_\mathrm{cavity}) = \left(0.5 ~ \frac{R_*}{R_\mathrm{cavity}}\right)^{\frac{2}{4+\beta_\mathrm{opacity}}} \times T_*,
\end{equation}
where $\beta_\mathrm{opacity}$ represents the exponent of the slope of the frequency-dependent opacities at longer wavelengths.
The assumption of gray absorption would, e.g., imply $\beta_\mathrm{opacity} = 0$ and, therefore, results in a temperature slope of $T \propto R_\mathrm{cavity}^{-0.5}$.
The slope for the opacities of \citet{Laor:1993p736} is $\beta_\mathrm{opacity} \approx 2.05$.

We can express the radiative acceleration in the FLD case using Eq.~\ref{eq:RadiativeAcceleration} with the Rosseland mean opacities of Fig.~\ref{fig:OpacitiesFLD} 
at the local radiation temperature $T(R_\mathrm{cavity})$ computed via Eq.~\ref{eq:Spitzer}.
The radiative acceleration in the RT case is computed by using Eq.~\ref{eq:RadiativeAcceleration} with the corresponding Planck mean opacities (including the local dust evaporation probability) of Fig.~\ref{fig:OpacitiesRT} at the stellar temperature $T_*$, which is for a specific stellar mass $M_*$ also taken from the evolutionary track of \citet{Hosokawa:2009p12591}.
The final formulae for the accelerations are given by
\begin{eqnarray}
\label{eq:Acceleration-FLD}
a_\mathrm{rad}^\mathrm{FLD} &=& \kappa_\mathrm{R}(T(R_\mathrm{cavity})) \times \frac{L_*(M_*)}{4\pi~c~R_\mathrm{cavity}^2}, \\
\label{eq:Acceleration-RT}
a_\mathrm{rad}^\mathrm{RT} &=& \kappa_\mathrm{P}(T_*(M_*)) \times \frac{L_*(M_*)}{4\pi~c~R_\mathrm{cavity}^2}, \\
\label{eq:Acceleration-Gravity}
a_\mathrm{grav} &=& \frac{G~M_*}{R_\mathrm{cavity}^2}.
\end{eqnarray}
The two remaining independent parameters are the actual stellar mass $M_*$ and the location $R_\mathrm{cavity}$ of the cavity shell material.
In the RT case, the opacity in regions cooler than the dust evaporation temperature is independent of the location $R_\mathrm{cavity}$, i.e.~the radiative acceleration scales with $a \propto R_\mathrm{cavity}^{-2}$ in the same way as the gravitational acceleration.
The ratio $a_\mathrm{rad}^\mathrm{RT} / a_\mathrm{grav}$ of these accelerations, therefore, is scale-free and depends only on the proto-stellar properties leading to the well-known formula for the generalized Eddington limit
\begin{equation}
\frac{L_*}{M_*} = \frac{4 \pi ~ G ~ c}{\kappa_\mathrm{P}(T_*)}
\end{equation}

\subsection{Results}
\begin{figure*}[htbp]
\begin{center}
\includegraphics[width=0.49\textwidth]{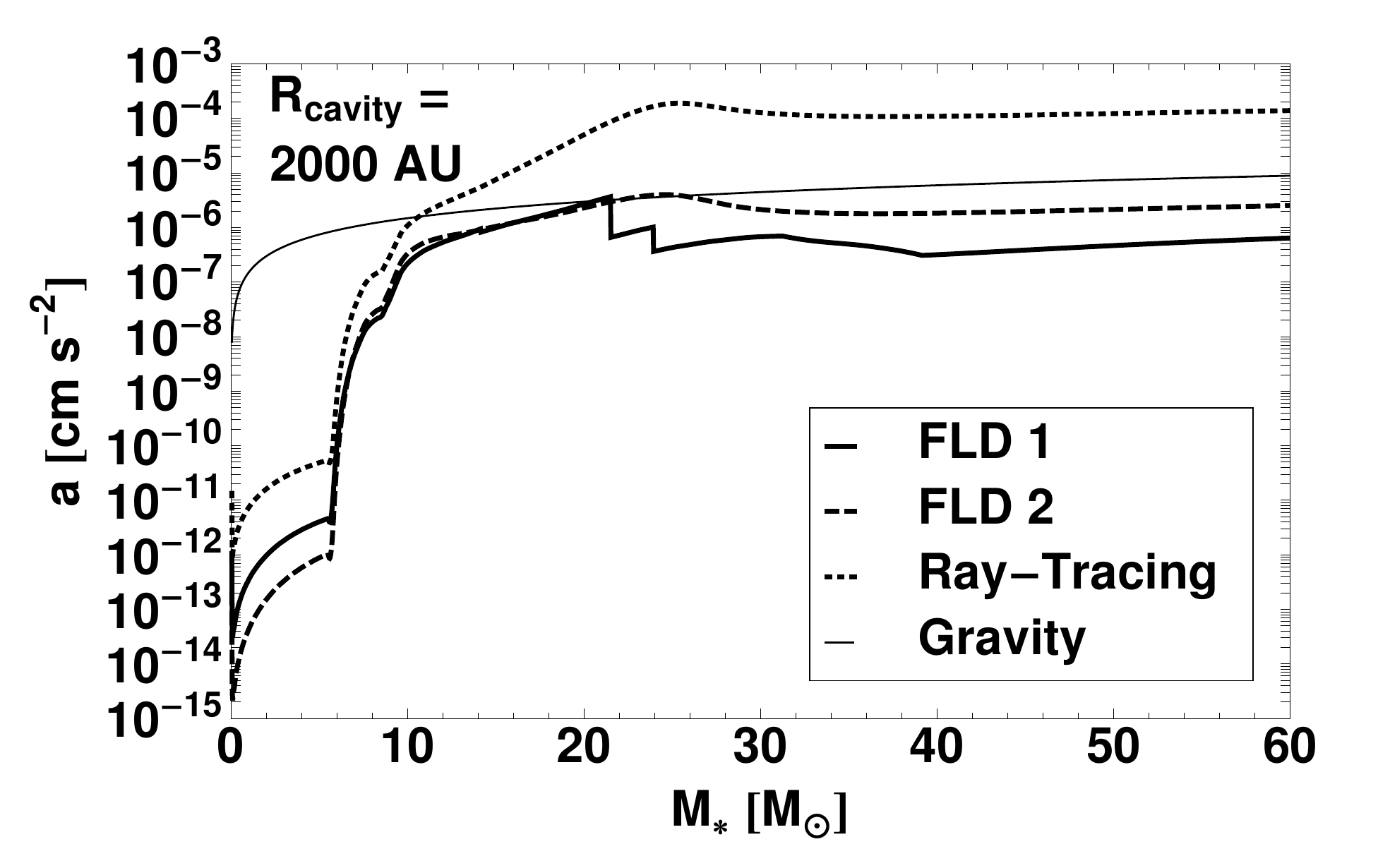}
\includegraphics[width=0.49\textwidth]{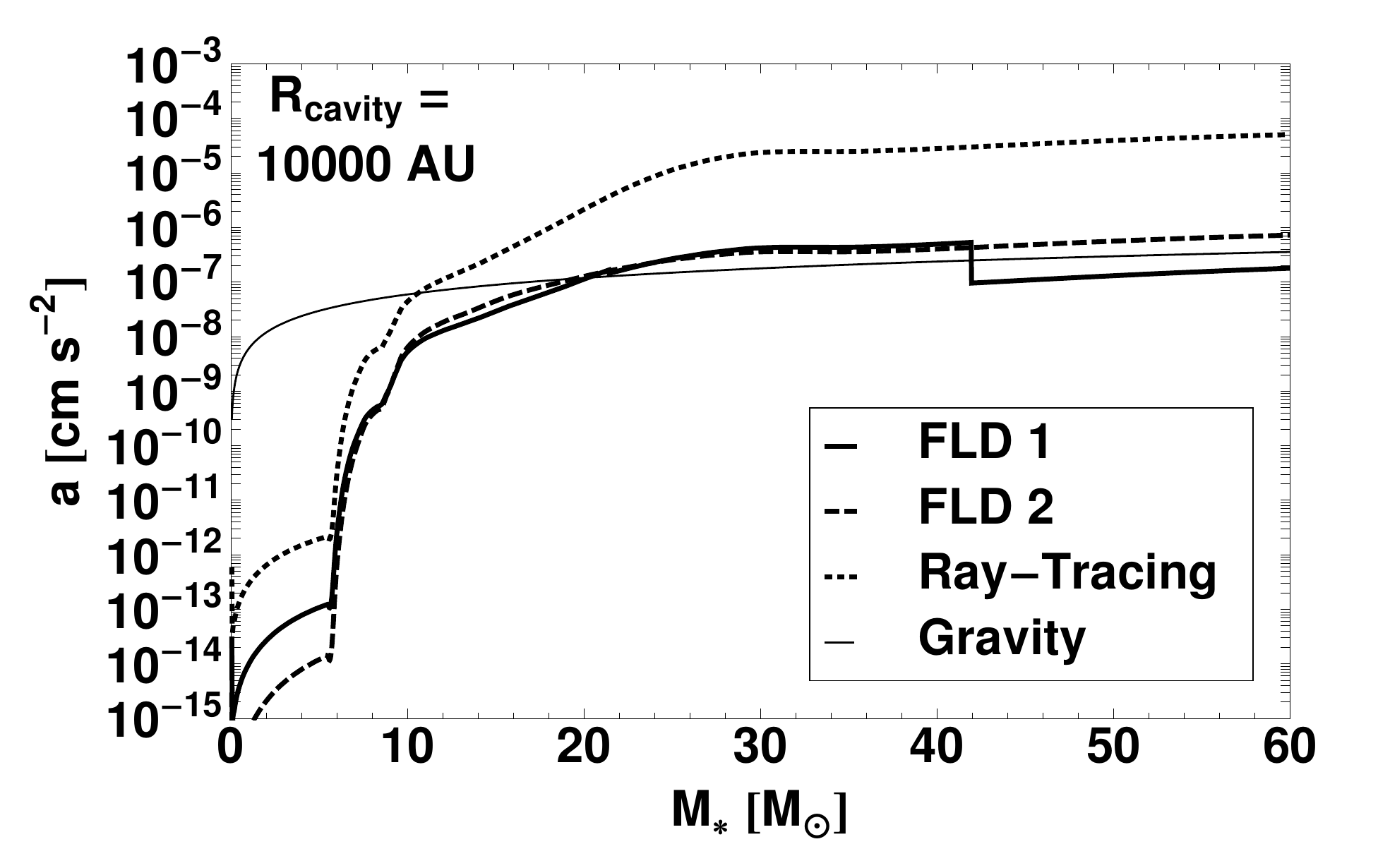}
\caption{
Radiative and gravitational acceleration at a position 
$R_\mathrm{cavity}=2000$~AU (left panel) and 
$R_\mathrm{cavity}=10000$~AU (right panel) 
above the massive proto-star as a function of the stellar mass $M_*$ for three different radiation transport approaches:
The label ``FLD 1'' denotes the FLD approximation with the opacity description of \citet{Krumholz:2007p1380}.
The label ``FLD 2'' refers to the FLD approximation with the opacities used in \citet{Kuiper:2010p17191, Kuiper:2011p17433} and in this paper.
The label ``Ray-Tracing'' refers to the RT step as in the hybrid radiation transport scheme.
}
\label{fig:AccelerationVSStellarMass}
\end{center}
\end{figure*}

In the following, we compare the accelerations depending on the remaining variables $M_*$ and $R_\mathrm{cavity}$.
Therefore, Fig.~\ref{fig:AccelerationVSStellarMass} compares the radiative accelerations in both the FLD approximation 
and the RT approach to gravity at two different locations from the proto-star as a function of the stellar mass $M_*$.
The steep increase in the radiative forces at roughly 6-8\Msol marks the point in time at which the stellar luminosity in the \citet{Hosokawa:2009p12591} tracks starts to dominate over the accretion luminosity.

The results of the analytical estimate derived in the previous section and visualized in Fig.~\ref{fig:AccelerationVSStellarMass} fully support the outcome of the radiation hydrodynamical simulations:
In the RT case, the cavity shell becomes super-Eddington (ratio of radiative to gravitational acceleration of unity) for a 10\Msol star and beyond.
Afterwards, the radiative acceleration increases rapidly to be one or two orders of magnitude higher than gravity.
The cavity shell is highly super-Eddington in all RT+FLD simulations.
Furthermore, the continuous enhancement in radiative acceleration from a 10~\Msol up to a 26~\Msol star explains the difference between the first and second outflow launch in the low-mass ($M_\mathrm{core}=50\Msol$) case, which are not detected in the $M_\mathrm{core}=100\Msol$ cases, yielding higher accretion rates and more rapid stellar evolution.

At large radii, also in the FLD runs, the cavity shell is finally super-Eddington, but even for a 30\Msol star only by a factor of a few.
Most of the time, the cavity shell in the FLD case is in marginal Eddington equilibrium.
In the runs for the FLD approximation, the cavity shell can be estimated to be in marginal Eddington equilibrium for a proto-stellar mass of roughly 20\Msol and an extent of the cavity of roughly 2000~AU (left panel of Fig.~\ref{fig:AccelerationVSStellarMass}).
As a consequence, the launched cavity shell stops its expansion along the polar axis at a height $R_\mathrm{cavity} = 2000$~AU, where it undergoes an epoch of marginal Eddington equilibrium, i.e.~radiative forces are balanced by gravity.
In this situation, these cavity shells are supposed to become radiative Rayleigh-Taylor unstable \citep{Jacquet:2011p18452}.
This is in full agreement with the FLD simulation results herein.
 
This behavior changes drastically when the direct absorption of stellar light is taken into account.
In the simulations, which treat the stellar irradiation with a RT step, the dust grains of the cavity shell absorb the stellar light 
\vONE{according to}
its frequency.
In this way, the acceleration of the shell turns out to be one to two orders of magnitude higher than in the FLD approximation.
In the analytical estimate of the RT case, the environment of the 20\Msol proto-star is super-Eddington by a factor of 20 and increases at large radii up to 100 for a proto-star of 30\Msol.

In the simulations, the difference in the radiation transport methods exists only up to an optical depth of $\tau(\nu) \approx 1$ in the cavity shell, i.e.~up to the location where all stellar photons are absorbed.
In our hybrid radiation transport scheme, the re-emission of the photons by dust grains (mostly at longer wavelengths in the infrared regime) is computed in the FLD solver step.  
The corresponding length scale is given by $l_* = (\kappa_*(\nu) ~ \rho_\mathrm{cavity})^{-1}$ and therefore depends on the frequency $\nu$ of the photons as well as on the actual shell density $\rho_\mathrm{cavity}$.
This absorption length scale $l_*$ is shown as a function of frequency $\nu$ for different densities $\rho$ of the cavity shell in Fig.~\ref{fig:Resolution}.

\section{Comparisons}
\label{sect:Comparisons}
\subsection{Comparison with a 3D gray FLD simulation}
\label{sect:Krumholz}
\citet{Krumholz:2009p10975} presented a self-gravity radiation hydrodynamics simulation of the collapse of a 100\Msol pre-stellar core.
The outer core radius was chosen to be 0.1~pc and the density profile declines in proportion to $r^{-1.5}$.
The initial isothermal temperature of the core is 20~K.
The pre-stellar core is initially in solid-body rotation without any turbulent motion.
The model describes a so-called monolithic collapse scenario.
The applied radiation transport method is the gray flux-limited fiffusion approximation.
These properties of this configuration are the same as in our simulation run ``C-FLD''.
The equations are solved on a 3D adaptive mesh refinement grid in cartesian coordinates.
Densities above the Jeans density on the finest grid level are represented by sink particles.

During the simulation, bipolar ``radiation-filled bubbles'' are blown into the environment of the forming massive star.
\vONE{To clarify our}
terminology, these ``radiation-filled bubbles'' correspond to the ``radiation-pressure-dominated cavities'' and the ``bubble wall'' is called a ``cavity shell'' in this paper.
At an extent of roughly from 1200 to 2000~AU (from \citet{Krumholz:2009p10975}, Fig.~1 therein) the cavity shell is subject to a ``radiative Rayleigh-Taylor instability'', meaning that the radiation pressure expands the optically thin region only at specific solid angles, while at other solid angles the concentrated mass load on top of the cavity shell is able to penetrate into the low-density region.

This mechanism of shell instability, the focus of radiation pressure onto solid angles with lower optical depth, and the collapse of condensed material from the infalling envelope at other solid angles, precisely matches the outcome of our simulations, if the FLD approximation is used for the stellar radiation feedback.
After the epoch of instability -- the caving-in of material -- the continuing evolution of their simulation differs in terms of morphology from our simulations because of the evolving non-axisymmetric structures.
We comment further on this difference after the following paragraph on the RT + FLD simulations.

In our simulations using the frequency-dependent RT step for the stellar irradiation, the mass load on top of the cavity shell causes a dependence of the optical depth on the solid angle (owing to the decreasing centrifugal forces towards the pole).
However, in these simulations, the caving-in of material at solid angles of high optical depth is prohibited; 
this is most likely caused by the treatment of the direct stellar irradiation by means of a RT approach preserving the natural isotropy of the irradiation up to the first absorption layer.
Furthermore, the RT approach accounts for the respective frequency-dependent absorption coefficients as demonstrated in the previous sections. 

\vONE{The cavity is affected by Rayleigh-Taylor-like instabilities, when the FLD approximation is used for the radiation transport. 
This implication is verified by the axial symmetric simulations performed herein as well as by the 3D simulation in \citet{Krumholz:2009p10975}. 
Improving the radiation transport scheme by incorporating a ray-tracing of the direct stellar irradiation corrects for this behavior and leads to a stable cavity growth. 
This implication is verified by the axial symmetric simulations performed herein as well as by the 3D simulation in \citet{Kuiper:2011p17433}.
}
The largest differences between the numerical treatment in the simulation of \citet{Krumholz:2009p10975} and the simulations with similar initial conditions presented herein are 
\vONE{twofold:}
the radiation transport approaches and 
our 
\vONE{assumed}
axial symmetry. 
As mentioned above, the simplification to axial symmetry increases the difficulty of a comparison of epochs after the occurrence of the instability, but the overall simulation results indicate that the outcome, regardless of whether the instability occurs, does not depend on the axial symmetric approach:
\begin{itemize}
\item The 2D simulations in the FLD approximation match the outcome of the 3D simulation in the FLD approximation, namely the instability of the cavity.
\item In the 2D simulations with the improved radiation transport scheme (ray-tracing + FLD), the cavity remains stable.
\item In our 3D simulation \citep{Kuiper:2011p17433}, using a frequency-dependent RT step to account for stellar irradiation, the outflow cavity contains strong non-axisymmetric features, but no instability is detected.
\end{itemize}
This implies that the stability of the outflow cavity is strongly effected by the direct stellar irradiation of the massive star, independent of the symmetry.

\subsection{Comparison with frequency-dependent FLD simulations}
In \citet{Yorke:2002p1}, the authors presented six self-gravity radiation hydrodynamics simulations of the collapse of a pre-stellar core.
The initial core mass in the simulations was chosen to be 30, 60, and 120\Msol with an outer core radius of 0.05, 0.1, and 0.2~pc.
The density profile drops proportional to $r^{-2}$.
Since in these simulations an additional inflow of mass at the outer boundary into the computational domain is allowed, the total mass reservoir of the forming star is not limited to these initial core masses.
The initial isothermal temperature of the core is 20~K.
The pre-stellar core is initially in solid-body rotation without any turbulent motion.
The equations are solved on a static nested grid of three levels in cylindrical coordinates assuming axial symmetry and midplane symmetry.
The forming massive star is represented by a sink cell at the origin of the domain.
The radiation transport method used is the gray as well as the frequency-dependent FLD approximation.

The outflow properties in the simulations with gray FLD and initial core masses of 30 and 120\Msol are neither visualized nor discussed in \citet{Yorke:2002p1}.
In the simulation with gray FLD and an initial core mass of 60\Msol, the massive star stops its mass growth at $M_* = 20.7\Msol$ and maintains a luminosity of $L_* = 5.2 \times 10^4 \Lsol$.
No polar cavity forms before the end of the simulation at 110~kyr.

In the three simulations with frequency-dependent FLD, polar cavities are formed.
In the 30\Msol case, the cavity interacts with the infalling envelope and collapses within a few thousand years.
In the two higher mass cases, the infall even at larger radii is reversed shortly after the launch of the outflow cavity; therefore the cavity does not interact with a gravitationally driven infall anymore and simply expands in time.
The 60\Msol simulation run was stopped after an evolutionary age of 45~kyr.
Nevertheless, in the 120\Msol case, the radiation pressure 
\vONE{eventually}
evacuates the stellar environment in all directions, including the disk midplane.
After its first expansion phase, this radiation-pressure-dominated cocoon-like structure decreases again, especially in the direction of the highest optical depth (the midplane), similar to the behavior of the expansion of the cavity shell in the FLD simulation by \citet{Krumholz:2009p10975} as well as in our FLD simulations.
The expansion velocity of the cavities of $v \gtrsim 10 \mbox{ km s}^{-1}$ corresponds more to the velocities observed in our FLD simulation ($v = 6-30 \mbox{ km s}^{-1}$) than those using the ray-tracing method ($v \sim 100 \mbox{ km s}^{-1}$).

The frequency dependence of the FLD solver (as in our RT method) might also be important.
Using the FLD approximation in the gray limit leads to a downgrading of the radiation field throughout the cavity region, i.e.~the dust grains in the cavity shell absorb the stellar flux with the Rosseland mean opacity at the local radiation temperature of the cavity shell, which does not resemble the broad stellar irradiation spectrum.
By dividing the stellar flux into several frequency bins and treating the flux of each bin in the FLD approximation, one also can ensure that the high-frequency part of the stellar spectrum is absorbed with the appropriate absorption coefficient.
\vONE{In addition to a more correct description of the}
absorption behavior in the cavity shell, the RT approach provides a most realistic representation of the photon paths emitted at the stellar photosphere.
Owing to the integral over the solid angle in the FLD approximation, the resulting flux does not include the correct propagation direction.

Moreover, including the long-wavelength part of the stellar spectrum explicitly can accelerate the layers on top of the swept-up cavity shell, which is optically thick for the short wavelengths only.
Furthermore, \citet{Schartmann:2011p17754} demonstrated \vONE{that} in numerical simulations using a RT approach for the radiation feedback, the acceleration of a gas clump by radiation pressure against a gravitational potential is inversely proportional to the optical depth of the clump, as expected by analytic estimates similar to the computation of the Eddington limit.

\subsection{Comparison with an analytic stability analysis}
\citet{Jacquet:2011p18452} derived a criterion for the stability of radiation-pressure-dominated cavities for the adiabatic approximation.
They provided growth times of the radiative Rayleigh-Taylor instability in cavity shells around massive stars.

This analytic analysis fully supports the outcome of our investigations.
From our estimate of the accelerations in Sect.~\ref{sect:Analytic}, we compute the expansion timescale of the cavity shell to be
\begin{equation}
t_\mathrm{cavity} = \sqrt{\frac{2~R_\mathrm{cavity}}{a_\mathrm{rad} - a_\mathrm{grav}}}.
\end{equation}
In the case of FLD simulations, the radiation pressure onto the cavity shell is in marginal equilibrium with gravity ($E \approx 1$) for long epochs in time.
The corresponding shell expansion timescale goes to infinity during these epochs, allowing the radiative Rayleigh-Taylor instability to set in.
In the case of RT simulations, radiation pressure exceeds gravity by 1-2 orders of magnitude, leading to a much shorter cavity shell expansion timescale.
For a stellar mass of 20\Msol and a cavity radius of $R_\mathrm{cavity} = 1400$~AU, one obtains an expansion timescale of $t_\mathrm{cavity} \approx 0.68$~kyr.

\citet{Jacquet:2011p18452} stated that the growth times of the radiative Rayleigh-Taylor instability in cavity shells around massive stars should always be shorter then 1~kyr, but actually the numbers one reads from their plots 
\vONE{(Fig.~3 and 4)} 
are instead $6.3 - 40$~kyr depending on the actual stellar mass and the wavelength of the instability. 
Hence, the expansion timescale of the shell is at least one order of magnitude shorter than the timescale for the growth of the radiative Rayleigh-Taylor instability.
Even, for the largest shell radius of $R_\mathrm{cavity} = 0.1$~pc at the outer boundary of our computational domain, the expansion timescale is $t_\mathrm{cavity} \approx 10$~kyr.
During this 10~kyr, the massive star increases its mass approximately from 20 to 30\Msol by disk accretion. 
Therefore, a cavity shell that  has formed around a massive star is expected to be able to expand up to the stellar cluster scale ($\approx 0.1$~pc) before being prone to the radiative Rayleigh-Taylor instability.

One might discuss whether an actual outflow of a massive star might fulfill the conditions for adiabaticity.
In \citet{Jacquet:2011p18452}, the authors assume that the cavity shell is in the ``adiabatic regime''.
This is estimated in \citet{Jacquet:2011p18452} in Eq.~(84) under the assumption of a hot cavity shell with $T \lesssim 1100$~K at a very large radius of $R_\mathrm{cavity} \approx 10^4$~AU.
In the simulation data of \citet{Krumholz:2009p10975} and our simulations in this paper, we observe much lower shell temperatures, even at much smaller radii.
As a consequence, for the stellar evolutionary tracks of \citet{Hosokawa:2009p12591} and the shell temperature estimate given in Eq.~\eqref{eq:Spitzer}, the cavity shell is in the non-adiabatic regime even out to an extent of 0.1~pc for a stellar mass of $M_* \le 40 \mbox{ M}_\odot$.
In the non-adiabatic regime, the radiative Rayleigh-Taylor instability is damped by diffusion.
The quantitative change caused by the damping remains unclear, and the growth rates of the radiative Rayleigh-Taylor instability in the non-adiabatic regime have not yet been derived.

\subsection{Comparison with observations}
\label{sect:Observations}
As mentioned in the introduction, radiation-pressure-dominated cleared cavities or optically thin bubbles of the discussed sizes around massive proto-stars 
\vONE{have not yet been}
observed.
Similar cavities are also proposed to form during a non-radiative but magnetized pre-stellar core collapse as shown in MHD simulations of \citet{Banerjee:2006p13404, Banerjee:2007p13399} and \citet{Hennebelle:2011p17895}.
The observation of cavities, especially in their initial launch phase, are hampered by the extinction (the cavities are still embedded in the infalling envelope) as well as the limited resolution of the inner core structure close to the proto-star.
\vONE{However,}
large-scale polar cavities could 
\vONE{potentially} 
diminish the extinction of the envelope at pole-on viewing angles.

Moreover, a direct comparison of the detailed morphology of the simulated outflow cavities is of course restricted by the limitations and caveats of the numerical simulations discussed in Sect.~\ref{sect:Limitations}.
Nevertheless, the basic properties of the radiation-pressure-dominated outflows investigated in this paper include a shell velocity on the order of $v \approx 100 \mbox{ km s}^{-1}$, as well as a mass-loss rate of roughly $\dot{M} \approx 10^{-4} \Msol \mbox{ yr}^{-1}$ (see Fig.~\ref{fig:1Danalysis}).

\section{Summary}
\label{sect:Summary}
We have performed six self-gravity radiation hydrodynamics simulations of a collapsing pre-stellar core including the launch of a radiation-pressure-dominated outflow cavity by the centrally forming star.
For the radiation transport method, 
three of these simulations were performed using the FLD approximation and
three use the hybrid radiation transport scheme introduced in \citet{Kuiper:2010p12874}.
In the hybrid radiation transport scheme, the stellar irradiation is computed in a frequency-dependent RT step and only the computation of the thermal dust emission makes use of the FLD approximation.
The three different initial conditions of the simulations vary in terms of the initial pre-stellar core mass (50 and 100~$\mbox{M}_\odot$) as well as the initial density slope ($\rho \propto r^{-1.5}$ and $\rho \propto r^{-2}$).
The analysis of these simulations indicates that only in the case of the FLD approximation does the cavity shell undergo an epoch of marginal Eddington equilibrium, i.e.~radiative forces are balanced by gravity.
As a consequence, the expansion of the outflow cavity along the polar axis stops and the infalling envelope material gathered on top of the cavity shell caves in.
In contrast to this scenario, simulations using the hybrid radiation transport scheme predict that there are stable and rapidly growing outflow cavities, and that no long-term epochs of marginal Eddington equilibrium occur.

We analyzed the outcome of these simulations in more detail using analytical estimates of the radiative acceleration and the growth timescale of the cavities with respect to the radiation transport method.
In contrast to the FLD assumption that dust grains absorb photons according to the local radiation temperature, dust grains in the shell on top of the optically thin cavity are directly irradiated by photons from the luminous massive star.
Therefore, including the feedback effect of direct stellar irradiation via a frequency-dependent RT approach leads to the prediction of stronger radiative forces in the cavity shell.
Including the direct absorption of stellar photons enhances the acceleration of the shell by between one and two orders of magnitude in comparison with the FLD assumption.
This analytic estimate confirms the much higher shell velocity detected in the radiation hydrodynamics simulations including the RT step for stellar radiation feedback.
These estimates of the relevant timescales imply that this enhancement in acceleration leads to a cavity expansion time that is much shorter than the timescale needed for the radiative Rayleigh-Taylor instability to occur.

The results of this study does not prohibit in general the existence of (radiative) instabilities in cavity shells.
The simulation results for the three different initial conditions show a dependence of the radiative launching and cavity growth on the density distribution of the stellar environment, e.g.~initially flatter density profiles lead to larger amounts of mass at larger radii and, therefore, a heavier mass load on top of the cavity shell.
The limitations and caveats of the numerical simulations (e.g.~the assumptions of no magnetic fields and perfect dust-to-gas coupling) do not allow a complete or even comprehensive description of the complex outflow region of a massive star.

Nevertheless, this study clearly emphasizes the strong influence of the (spectral) stellar irradiation on both the dynamics and the stability of the first absorption layer. 
It has been demonstrated that a careful numerical treatment of this direct stellar irradiation feedback can be achieved by using a RT approach.
For comparison with future high-resolution observations of the inner core regions, the velocity and the mass-loss rate of these radiation-pressure-dominated outflows are given: in the simulations, \vONE{we find} a shell velocity on the order of $v \approx 100 \mbox{ km s}^{-1}$ and a mass-loss rate of roughly $\dot{M} \approx 10^{-4} \Msol \mbox{ yr}^{-1}$.
The qualitative and quantitative analyses of our numerical simulations,
including different initial conditions and radiation transport methods
combined with the analytical estimates of the relevant forces and timescales,
cast severe doubt on the occurrence of the radiative Rayleigh-Taylor instability in the radiation-pressure-dominated cavities around forming high-mass stars.

\begin{acknowledgements}
We thank Malcolm Walmsley and Harold Yorke for fruitful discussions of the paper.
We acknowledge the help of Andrea Mignone, the main developer of the open source MHD code Pluto, as well as Petros Tzeferacos, who implemented the viscosity tensor into Pluto.
We thank our colleague Mario Flock for his dedicated participation in the enhancements of our code version and Takashi Hosokawa for contributing the stellar evolutionary tracks.
Thanks to Sarah Ragan for proof-reading the manuscript.
\end{acknowledgements}

\bibliographystyle{aa}
\bibliography{Papers}

\end{document}